\documentclass[12pt,a4]{article}
 \usepackage{amsmath}
 \usepackage{mathtools} 
 \usepackage{graphicx}
 \usepackage[merge,numbers,compress]{natbib}
 \usepackage[T1]{fontenc}
 \usepackage{booktabs}
 \usepackage{xcolor} 
 \usepackage{xspace}
 \usepackage{dcolumn}
 \usepackage{placeins}
 \usepackage[colorlinks=true,citecolor=blue!50!black,linkcolor=black]{hyperref}
 \usepackage{caption}
 \usepackage
 [subrefformat=parens,position=top,skip=-15pt,margin=15pt,justification=justified,singlelinecheck=false]
 {subcaption}
 \usepackage{multirow}
\usepackage{slashed}
\usepackage[bottom]{footmisc}
\usepackage{comment}
\usepackage{bigints}
\usepackage{hhline}
\usepackage[normalem]{ulem}

\newcommand{\processmue}{\gamma\gamma \to \tau^+ \tau^- \to e^+ \mu^- \bar{\nu}_\tau \nu_\tau \bar{\nu}_{\mu} \nu_{e}}

\newcommand{\process}{\gamma\gamma \to \tau^+ \tau^- \to \ell'^+ \ell^- \bar{\nu}_\tau \nu_\tau \bar{\nu}_{\ell} \nu_{\ell'}}

\newcommand{\M}{\mathcal{M}}

\def\EXP#1{{\ensuremath \times 10^{#1}}}
\newcommand{\LO}{\ensuremath{\text{LO}}}
\newcommand{\NLO}{\ensuremath{\text{NLO}}}

\newcommand{\ri}{\mathrm i}

\newcommand{\ie}{\emph{i.e.}\ }
\newcommand{\eg}{\emph{e.g.}\ }

\let\origfootnote\footnote
\renewcommand{\footnote}[1]{\kern.06em\origfootnote{#1}}
\newcommand{\punctfootnote}[1]{\kern-.06em\origfootnote{#1}}

\def\be{\begin{equation}}
\def\ee{\end{equation}}

\newcommand{\PH}{\ensuremath{\text{H}}\xspace}

\newcommand{\Pe}{\ensuremath{\text{e}}\xspace}
\newcommand{\Pb}{\ensuremath{\text{b}}\xspace}

\newcommand{\Pt}{\ensuremath{\text{t}}\xspace}
\newcommand{\Pu}{\ensuremath{\text{u}}\xspace}
\newcommand{\Pd}{\ensuremath{\text{d}}\xspace}
\newcommand{\Ps}{\ensuremath{\text{s}}\xspace}
\newcommand{\Pc}{\ensuremath{\text{c}}\xspace}

\newcommand{\PW}{\ensuremath{\text{W}}\xspace}
\newcommand{\PZ}{\ensuremath{\text{Z}}\xspace}

\newcommand{\ME}{\ensuremath{m_\Pe}\xspace}
\newcommand{\MM}{\ensuremath{m_{\mu}}\xspace}
\newcommand{\ML}{\ensuremath{m_{\tau}}\xspace}
\newcommand{\Mt}{\ensuremath{m_\Pt}\xspace}
\newcommand{\MH}{\ensuremath{M_\PH}\xspace}
\newcommand{\MWOS}{\ensuremath{M_\PW^\text{OS}}\xspace}
\newcommand{\MW}{\ensuremath{M_\PW}\xspace}
\newcommand{\MZOS}{\ensuremath{M_\PZ^\text{OS}}\xspace}
\newcommand{\MZ}{\ensuremath{M_\PZ}\xspace}
\newcommand{\Mb}{\ensuremath{m_\Pb}\xspace}
\newcommand{\Mc}{\ensuremath{m_\Pc}\xspace}

\newcommand{\GZOS}{\ensuremath{\Gamma_\PZ^\text{OS}}\xspace}

\newcommand{\GWOS}{\ensuremath{\Gamma_\PW^\text{OS}}\xspace}

\newcommand{\keV}{\ensuremath{\,\text{keV}}\xspace}

\newcommand{\GeV}{\ensuremath{\,\text{GeV}}\xspace}
\newcommand{\TeV}{\ensuremath{\,\text{TeV}}\xspace}

\newcommand{\order}[1]{\ensuremath{\mathcal{O}{\left(#1\right)}}\xspace}

\newcommand{\GF}{\ensuremath{G_\mu}}

\newcommand{\ptsub}[1]{\ensuremath{p_{\text{T},#1}}\xspace}

\renewcommand{\Re}{\mathop{\mathrm{Re}}\nolimits}

\newcommand{\MVOS}{\ensuremath{M_{\text{V}}^\text{OS}}\xspace}%
\newcommand{\GVOS}{\ensuremath{\Gamma_{\text{V}}^\text{OS}}\xspace}%

\newcommand{\vh}{\ensuremath{\vphantom{\int_A^A}}}

\newcommand{\newc}{\newcommand}
\newc{\bi}{\begin{itemize}}
\newc{\ei}{\end{itemize}}
\newc{\benu}{\begin{enumerate}}
\newc{\eenu}{\end{enumerate}}
\newc{\bc}{\begin{center}}
\newc{\ec}{\end{center}}
\newc{\bfig}{\begin{figure}}
\newc{\efig}{\end{figure}}
\newc{\qbar}{\bar{q}}
\newc{\go}{\tilde{g}}
\newc{\PB}{\textsc{Powheg-Box}}

\newcommand{\MonteTau}{{\sc MonteTau}\xspace}
\newcommand{\gammaUPC}{{\sc gamma-UPC}\xspace}
\newcommand{\libepa}{{\sc libepa}\xspace}
\newcommand{\TAUOLA}{{\sc TAUOLA}\xspace}
\newcommand{\PHOTOS}{{\sc PHOTOS}\xspace}
\newcommand{\Herwig}{{\sc Herwig}\xspace}
\newcommand{\cpp}{C\texttt{++}\xspace}
\newcommand{\VEGAS}{VEGAS\xspace}
\newcommand{\gsl}{GNU Scientific Library\xspace}
\newcommand{\Mathematica}{{\sc Mathematica}\xspace}
\newcommand{\FeynArts}{{\sc FeynArts}\xspace}
\newcommand{\FormCalc}{{\sc FormCalc}\xspace}
\newcommand{\McMule}{{\sc McMule}\xspace}
\newcommand{\Recola}{{\sc Recola}\xspace}
\newcommand{\Sherpa}{{\sc Sherpa}\xspace}

\newcommand{\collier}{{\sc Collier}\xspace}

\newcommand{\rT}{{\mathrm{T}}}
\newcolumntype{.}{D{.}{.}{-1}}
\newcolumntype{d}[1]{D{.}{.}{#1}}

\newcommand{\EW}{\ensuremath{\text{EW}}}

\newcommand{\bare}{{\ensuremath{\text{bare}}}}
\newcommand{\drs}{{\ensuremath{\text{drs}}}}

\colorlet{tableoverheadcolor}{gray!37.5}
\colorlet{tableheadcolor}{gray!25}
\colorlet{tablerowcolor}{gray!12.5}

\newcommand{\gsim}
{\;\raisebox{-.3em}{$\stackrel{\displaystyle >}{\sim}$}\;}
\def\asymp#1{\;\raisebox{-.4em}{$\widetilde{\scriptstyle #1}$}\;}

\newlength{\width}
\newlength{\height}

\def\draftdate{\relax}
\def\mda{\relax}
\def\mua{\relax}
\def\mla{\relax}
\def\draft{
\def\thtystars{******************************}
\def\sixtystars{\thtystars\thtystars}
\typeout{}
\typeout{\sixtystars**}
\typeout{* Draft mode!
         For final version remove \protect\draft\space in source file *}
\typeout{\sixtystars**}
\typeout{}
\def\draftdate{\today}
\def\mua{\marginpar[\boldmath\hfil$\uparrow$]%
                   {\boldmath$\uparrow$\hfil}\color{black}%
                    \typeout{marginpar: $\uparrow$}\ignorespaces}
\def\mda{\color{red}\marginpar[\boldmath\hfil$\downarrow$]%
                   {\boldmath$\downarrow$\hfil}%
                    \typeout{marginpar: $\downarrow$}\ignorespaces}
\def\mla{\marginpar[\boldmath\hfil$\rightarrow$]%
                   {\boldmath$\leftarrow $\hfil}%
                    \typeout{marginpar: $\leftrightarrow$}\ignorespaces}
\def\Mua{\marginpar[\boldmath\hfil$\Uparrow$]%
                   {\boldmath$\Uparrow$\hfil}\color{black}%
                    \typeout{marginpar: $\uparrow$}\ignorespaces}
\def\Mda{\color{red}\marginpar[\boldmath\hfil$\Downarrow$]%
                   {\boldmath$\Downarrow$\hfil}%
                    \typeout{marginpar: $\downarrow$}\ignorespaces}
\def\Mla{\marginpar[\boldmath\hfil\textcolor{red}{$\Rightarrow$}]%
                   {\boldmath\textcolor{red}{$\Leftarrow $}\hfil}%
                    \typeout{marginpar: $\leftrightarrow$}\ignorespaces}
\overfullrule 5pt
\oddsidemargin 15mm
\marginparwidth 29mm
}

\newcommand{\hl}{\vphantom{$\int_A^B$}}

\newcommand{\wm}{\phantom{$-$}}
\newcommand{\wn}{\phantom{$0$}}
\newcommand{\mwm}{\phantom{-}}

\newcolumntype{C}{>{\centering\arraybackslash}p{0.105\textwidth}}

\begin{document}

\title{\hfill ~\\[-30mm]
\phantom{h} \hfill\mbox{\small FR-PHENO-2025-004}
\\[1cm]
\vspace{13mm}   \textbf{Electroweak corrections to $\tau^+\tau^-$ production in ultraperipheral heavy-ion collisions at the LHC}}

\date{}
\author{
Stefan Dittmaier\footnote{E-mail:
  \texttt{stefan.dittmaier@physik.uni-freiburg.de}},
Tim Engel, 
José Luis Hernando Ariza\footnote{E-mail: 
  \texttt{jose.luis.hernando@physik.uni-freiburg.de}},
Mathieu Pellen\footnote{E-mail:
  \texttt{mathieu.pellen@physik.uni-freiburg.de}}
\\[9mm]
{\small\it Universit\"at Freiburg, Physikalisches Institut,} \\
{\small\it Hermann-Herder-Str.\ 3, 79104 Freiburg, Germany}\\[3mm]
}
\maketitle

\begin{abstract}
\noindent

While the anomalous magnetic moments of the electron and the muon have been
measured with remarkable precision, the magnetic moment of the $\tau$-lepton is only known to rather limited precision. 
A promising approach to measure it exploits $\tau^+\tau^-$ production in
ultraperipheral collisions of lead ions at the LHC. In this article,
a state-of-the-art theory prediction for $\tau^+\tau^-$
production including leptonic $\tau$-decays is provided. The impact of
spin correlations between the $\tau$-leptons, of the masses of final-state leptons, of next-to-leading-order electroweak corrections, and of the parametrization of the
photon flux are discussed.

\end{abstract}
\thispagestyle{empty}

\newpage
\setcounter{page}{1}

\tableofcontents

\section{Introduction}






The Standard Model of Particle Physics (SM) is, to date, the most successful theory for describing all known elementary particles and their fundamental interactions. 
Its validity has been tested with astonishing precision over the last decades, particularly at collider experiments.


Historically, an important type of precision observables used to test the SM comprises the anomalous magnetic moments of leptons, $a_{\ell}$. 
The anomalous magnetic moment of the electron, $a_e$, shows one of the best agreements observed in physics between theory~\cite{Kinoshita:2014lmy} and experiment~\cite{Rivas:2024eom}, matching up to 12 significant digits. 
In turn, the quantity $a_e$ provides one of the best ways to determine the fine-structure constant $\alpha(0)$.
On the other hand, the anomalous magnetic moment of the muon, $a_{\mu}$, has been a subject of significant discrepancy between theory~\cite{Aoyama:2020ynm} and experiments~\cite{Cotrozzi:2024bmc}, with a discrepancy rising up to $5.1\sigma$ in 2023.
The experimental determination of $a_\mu$ was performed by the Muon $g-2$ collaboration at the Brookhaven National Laboratory (BNL)~\cite{Muong-2:2006rrc} and at the Fermilab National Accelerator Laboratory (FNAL)~\cite{Muong-2:2021ojo,Muong-2:2023cdq}. 
The largest uncertainty on the theory side arises from the determination of the hadronic vacuum polarization, which was originally obtained using a data-driven dispersive approach based on electron--positron collisions.
Recently, lattice-based results~\cite{Toth:2022lsa} and a new $\tau$-data-driven analysis~\cite{Masjuan:2024ccq} have relaxed this tension to less than $2\sigma$. 
However, discrepancies with the $e^+e^-$-data-driven result~\cite{Aoyama:2020ynm} still need to be clarified.


In this context, studying the anomalous magnetic moment of the $\tau$-lepton, $a_{\tau}$, appears to be of major importance, especially since many Beyond Standard Model (BSM) theories predict a scaling of the BSM effects proportional to the mass squared of the considered lepton over the BSM scale squared, see \emph{e.g.}\ Ref.~\cite{Martin:2001st}. 
However, due to the small lifetime of the $\tau$-lepton, $\tau_{\tau} = 2.9\EXP{-13} \text{ s}$~\cite{ParticleDataGroup:2024cfk}, the experimental determination of $a_{\tau}$ constitutes a challenge. 
The first attempts to determine $a_{\tau}$ were made at the Large Electron--Positron Collider (LEP)~\cite{DELPHI:2003nah}. 
There, the process $e^+e^-\to e^+e^-\tau^+\tau^-$ was investigated, leading to the limits
\begin{align}\label{eq:DELPHI_limits}
  -0.052 < a_{\tau} < 0.013,  \qquad 95\%\text{ CL}.
\end{align}

Currently, the most precise determination of $a_{\tau}$ is obtained at the Large Hadron Collider (LHC) by the CMS collaboration~\cite{CMS:2024qjo} using proton--proton collisions. 
The resulting constraints on $a_{\tau}$ are
\begin{align}\label{eq:CMS_pp_limits}
  -0.0042 < a_{\tau} < 0.0062,  \qquad 95\%\text{ CL}.
\end{align}
However, these constraints are far away from the accuracy of the SM prediction~\cite{Keshavarzi:2019abf},
\begin{align}
  a^{\text{SM}}_{\tau} = (117717.1 \pm 3.9)\EXP{-8},
\end{align}
whose leading-order (LO) contribution is given by the so-called Schwinger term, 
$a^{(0),\text{SM}}_{\tau} = \alpha/(2\pi) = 0.00116\ldots$

In order to provide a more precise determination of $a_\tau$, the ATLAS~\cite{ATLAS:2022ryk} and CMS~\cite{Jofrehei:2022bwh} collaborations currently study $\tau$-pair production in ultraperipheral collisions (UPCs) of two lead ions, $\text{Pb}\text{Pb}\to\text{Pb}\text{Pb}\,\tau^+\tau^-$, at the LHC.
UPCs proceed via two nuclei passing each other at a separation larger than the sum of their radii, ensuring a large probability for the nuclei to remain intact in the interaction. 
The interaction itself occurs between the electromagnetic radiation emitted by the ions. 
Thus, the $\tau$-pair production is induced by light-by-light scattering, \ie $\gamma\gamma\to\tau^+\tau^-$. 
In comparison to inelastic proton--proton collisions, UPCs have the advantage of providing clean final states with extremely low background, as the nuclei do not break up. 
This approach builds on the same idea as $e^+e^-\to e^+e^-\tau^+\tau^-$ studies with the added benefit of a photon emission probability enhanced by a $Z^2$ factor for each ion, leading to a total $Z^4$ enhancement for the entire cross section, where $Z$ refers to the atomic number of the ion. 
Moreover, in comparison to other approaches to determine $a_{\tau}$, such as the studies of leptonic $\tau$-decays~\cite{Fael:2013ij} or of the process $e^+e^-\to\tau^+\tau^-$~\cite{Bodrov:2024wrw}, the UPCs are more sensitive to the $\gamma\tau\tau$ vertex due to its double appearance in the LO amplitude.
The sensitivity of UPCs to $a_{\tau}$ has been extensively discussed in the literature, see \emph{e.g.}\ Refs.~\cite{delAguila:1991rm,Atag:2010ja,Beresford:2019gww,Dyndal:2020yen,Verducci:2023cgx,Haisch:2023upo}.

Another interesting quantity accessible via UPCs, due to its charge-parity (CP) violating nature, is the electric dipole moment of the $\tau$-lepton, $d_{\tau}$. 
Within the SM, it arises only at three-loop level, leading to the suppression $\big|d_{\tau}^{\text{SM}} \big| < 10^{-34} e \,\text{cm}$~\cite{Hoogeveen:1990cb}. 
Currently, the best experimental limit has been obtained by CMS in proton--proton collisions~\cite{CMS:2024qjo},
\begin{align}\label{eq:CMS_pp_d_limits}
  \big|d_{\tau}\big| < 2.9\EXP{-17} e\,\text{cm}.
\end{align}

The production of a $\tau$-lepton pair in UPCs was first observed at the LHC by the CMS collaboration~\cite{Jofrehei:2022bwh} with the $\tau^-$-lepton decaying into a muon and the $\tau^+$-lepton decaying hadronically into 3 charged hadrons (3-prong decay). 
Afterwards, the ATLAS collaboration~\cite{ATLAS:2022ryk} confirmed this observation and further identified $\tau$-pair production via the signature where the $\tau^-$-lepton decays into a muon and the $\tau^+$-lepton decays either into one charged hadron or into a positron (1-prong decays). 
The limits on $a_{\tau}$ obtained from UPCs are, to date, competitive with those given in Eq.~(\ref{eq:DELPHI_limits}), but not with the constraint given in Eq.~(\ref{eq:CMS_pp_limits}).


On the theory side, the libraries \gammaUPC~\cite{Shao:2022cly} and \libepa~\cite{Zhemchugov:2023dfl} 
are state-of-the-art libraries to simulate the photon flux radiated off the heavy ions. 
Alternatively, UPCs can, for instance, be investigated with the public 
Monte Carlo generators SuperChic~\cite{Harland-Lang:2020veo} 
or Upcgen~\cite{Burmasov:2021phy}.
The process of $\tau$-pair production has been studied at next-to-leading-order (NLO) 
QED~\cite{Shao:2024dmk} and 
electroweak (EW) precision~\cite{Jiang:2024dhf,Shao:2025bma} considering only on-shell $\tau$-leptons. 
A simulation of the effect of a non-standard $a_{\tau}$ on the production of a $\tau$-lepton pair in UPCs including the $\tau$-lepton decays has also been 
studied in the literature~\cite{Dyndal:2020yen,Shao:2023bga}.
In many experimental analyses, $\tau$-lepton decays are treated with the Monte Carlo 
generator \TAUOLA~\cite{Jadach:1990mz} together with \PHOTOS~\cite{Barberio:1993qi} 
to simulate the QED radiative corrections to the corresponding decay. 
This implies that only final-state radiation is included in the description of the $\tau$-lepton decays, and the spin correlation between the $\tau$-leptons is neglected.
Spin correlations between $\tau$-leptons have been already studed in the literature for various processes using the Monte Carlo generators \Herwig~\cite{Bahr:2008pv,Bewick:2023tfi} and \Sherpa~\cite{Sherpa:2024mfk}.
These Monte Carlo generators include spin correlations via the algorithm presented in Ref.~\cite{Richardson:2001df}.
Finally, effects of dissociation or survival of the colliding hadrons on the determination
of $a_\tau$ and $d_\tau$ have been investigated in Ref.~\cite{Harland-Lang:2024zpn},
employing the {SuperChic} generator~\cite{Harland-Lang:2020veo,Harland-Lang:2021ysd}.


To achieve optimal SM predictions for the production of $\tau$-lepton pairs in UPCs, the decays of the $\tau$-leptons should both be included along with EW corrections.
In this work, predictions for $\tau$-pair production in UPCs are given considering two leptonically decaying $\tau$-leptons, \ie $\process$. 
The predictions are provided at NLO EW precision for $\ell=\mu$  and $\ell' = e$ retaining the masses of the final-state charged leptons and including spin correlations between the $\tau$-leptons.

The article is organised as follows: in Sec.~\ref{sec:features} the details of the calculation are described. 
Numerical results are presented in Sec.~\ref{sec:results}. 
In particular, the predictions for integrated and differential cross sections are given together with a discussion of the relevant effects included in the calculation: spin correlations between the $\tau$-leptons, masses of the final-state leptons, NLO EW corrections, and the influence of different parametrizations of the photon flux. 
Finally, a summary and concluding remarks are provided in Sec.~\ref{sec:conclusion}.

\section{Features of the calculation}
\label{sec:features}

In this section, we discuss the key aspects of our predictions for $\tau$-pair production via photon--photon scattering induced by UPCs of lead ions, assuming that the $\tau$-leptons decay leptonically. 
In Sec.~\ref{sec:EPA}, the equivalent-photon approximation (EPA) is introduced along with the theoretical framework employed to describe UPCs.
The hard process is described in Sec.~\ref{sec:hard_process}. 
Furthermore, the use of the narrow-width approximation (NWA) in the calculation is justified there, along with the difference between its naive and improved versions, where the latter includes spin correlations of the produced $\tau$-leptons.
In Sec.~\ref{sec:NLO}, a discussion of the NLO EW correction to the hard process within the NWA is provided.
Finally, the technical aspects and the set-up of the calculation are given in Secs.~\ref{sec:tools} and \ref{sec:setup}, respectively.

\subsection{Equivalent-photon approximation}
\label{sec:EPA}

\begin{figure}
    \centering{
    \raisebox{0pt}{\includegraphics[width=0.4\textwidth]{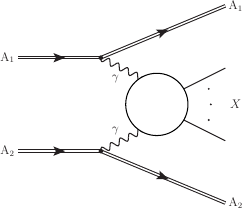}}
    }
    \caption{
    Structural diagram for ultraperipheral collisions of two ions $\mathrm{A}_1$ and $\mathrm{A}_2$. 
    The blob stands for any production mechanism of a final state $X$ via photon--photon scattering.
    }
    \label{fig:UPC}
\end{figure}

In order to describe the ultraperipheral collision of lead ions, we use the EPA~\cite{vonWeizsacker:1934nji,Williams:1934ad} 
which describes the electromagnetic field of any charged particle accelerated to high energies as a source of quasireal photons. 
In this approximation, the cross section $\sigma$ for the elastic production of a final state $X$ via photon--photon fusion is factorized in terms of the convolution of the so-called photon flux with the cross section $\hat{\sigma}$ of the hard process over the energies of the generated photons (see Fig.~\ref{fig:UPC}), \ie
\begin{align}
  \sigma(\sqrt{s_{\mathrm{A}_1\mathrm{A}_2}}) 
  = \int \frac{\mathrm{d} E_{\gamma_1}}{E_{\gamma_1}} 
         \frac{\mathrm{d} E_{\gamma_2}}{E_{\gamma_2}} 
         \frac{\mathrm{d}^2 N^{(\mathrm{A}_1\mathrm{A}_2)}_{\gamma_1/Z_1, \gamma_2/Z_2}}{\mathrm{d}E_{\gamma_1}\mathrm{d}E_{\gamma_2}} \,  
         \hat{\sigma}(\sqrt{s_{\gamma\gamma}}) ,
\end{align}
where $\mathrm{A}_1$ and $\mathrm{A}_2$ are ions with atomic numbers $Z_1$ and $Z_2$, respectively.
The factor 
$\frac{\mathrm{d}^2 N^{(\mathrm{A}_1\mathrm{A}_2)}_{\gamma_1/Z_1, \gamma_2/Z_2}}{\mathrm{d}E_{\gamma_1}\mathrm{d}E_{\gamma_2}}$ 
represents the photon flux, which gives the probability of a photon $\gamma_1$ being emitted by the ion $\mathrm{A}_1$ with an energy $E_{\gamma_1}$ and a photon $\gamma_2$ being emitted by the ion $\mathrm{A}_2$ with an energy $E_{\gamma_2}$, while the corresponding heavy ions do not break up.
While the cross section of UPCs is a function of the centre-of-mass energy 
$\sqrt{s_{\mathrm{A}_1\mathrm{A}_2}}$ 
of the two heavy ions, the cross section of the hard process is a function of the centre-of-mass energy 
$\sqrt{s_{\gamma\gamma}} = \sqrt{4E_{\gamma_1}E_{\gamma_2}}$ 
of the produced photons. 

The EPA allows for a small range of photon virtualities $Q_{\gamma_i}^2$, with an upper cut-off in the virtualities given by $Q_{\gamma_i}^2 < 1/R_{\mathrm{A}_i}^2$, where $R_{\mathrm{A}_i}$ is the charge radius of the ion $\mathrm{A}_i$~\cite{Budnev:1975poe}, 
\eg for lead ions $R_{\mathrm{Pb}}\approx 7\,\mathrm{fm}$ and thus $Q_{\gamma_i}^2 < 10^{-3}\GeV^2$.
However, for our calculation the photons are strictly treated as real, \ie $Q^2_{\gamma_i}=0$, as required for a consistent evaluation of the hard matrix element, ensuring gauge invariance. 
This treatment is justified by requiring that the heavy ions remain intact after the interaction.
As a consequence, we do not encounter initial-state corrections from photons radiated off quarks.
This assumption is also reflected in the choice of input-parameter scheme, because the photons couple effectively with $\alpha(0)$ in this case (see Sec.~\ref{sec:setup} for a more detailed discussion).

In the literature, mainly two approaches have been proposed to parametrize the photon flux for ultraperipheral heavy-ion collisions: One is based on the electric dipole form factor (EDFF) of the heavy ions~\cite{Cahn:1990jk}, another on the charge form factor (ChFF)~\cite{Vidovic:1992ik}. 
Both variants are supported by the \gammaUPC library~\cite{Shao:2022cly}, which 
is used in our calculation and also includes
a non-trival survival probability for the heavy ions to stay intact.
%
In the following, we focus on the description of the hard-scattering process.

\subsection{Hard process}
\label{sec:hard_process}

\begin{figure}
    \centering{
    \raisebox{0pt}{\includegraphics[width=0.24\textwidth]{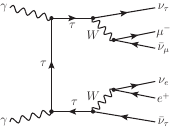}}
    \raisebox{0pt}{\includegraphics[width=0.24\textwidth]{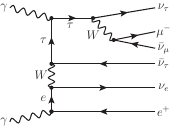}}
    \raisebox{0pt}{\includegraphics[width=0.24\textwidth]{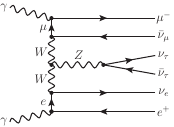}}
    \raisebox{0pt}{\includegraphics[width=0.24\textwidth]{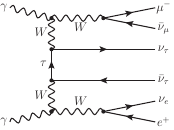}}
    \newline 
    (a) \hspace{3.2cm} (b) \hspace{3.2cm}  (c) \hspace{3.2cm} (d) \hspace{-2.9cm}
    \newline
    \raisebox{0pt}{\includegraphics[width=0.24\textwidth]{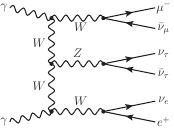}}
    \raisebox{0pt}{\includegraphics[width=0.24\textwidth]{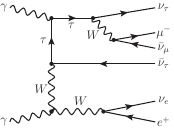}}
    \raisebox{0pt}{\includegraphics[width=0.24\textwidth]{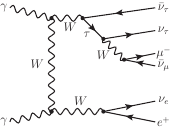}} 
    \raisebox{0pt}{\includegraphics[width=0.24\textwidth]{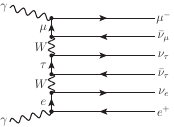}}
    \newline 
    (e) \hspace{3.2cm} (f) \hspace{3.2cm}  (g) \hspace{3.2cm} (h) \hspace{-1cm}
    }
    \caption{Examples of LO Feynman diagrams for the process $\gamma\gamma\to e^+\mu^-\bar{\nu}_\tau\nu_\tau\bar{\nu}_\mu\nu_{e}$: 
    (a) with two resonant $\tau$-leptons, 
    (b) with one resonant $\tau$-lepton, 
    (c) with one resonant gauge boson, 
    (d) with two resonant gauge bosons, 
    (e) with three resonant gauge bosons, 
    (f) with one resonant gauge boson and one resonant $\tau$-lepton,
    (g) with two resonant gauge bosons and one resonant $\tau$-lepton, 
    and (h) without resonances.}
    \label{fig:LO}
\end{figure}

The hard process of interest is the production of a $\tau$-pair via photon--photon scattering, assuming that both $\tau$-leptons decay leptonically, one into a positron and neutrinos, and the other into a muon and neutrinos, \ie 
\begin{align} \label{eq:process}
  \processmue.
\end{align}
In order to reach an appropriate treatment of the $\tau$-resonances, we start our discussion from the full off-shell process
$\gamma\gamma\to e^+\mu^-\bar{\nu}_\tau\nu_\tau\bar{\nu}_\mu\nu_e$
with any possible intermediate states.
In perturbation theory, the squared LO amplitude to the full off-shell process is of $\order{\alpha^6}$.
However, already at this order, there are more than hundred diagrams that produce the same final state via photon--photon scattering, see Fig.~\ref{fig:LO}, and they all contribute to the matrix element $\M$, including the subcontribution with two resonant $\tau$-leptons.



Among all diagram types shown in Fig.~\ref{fig:LO}, the type~(a) is by far dominating.
Each missing $\tau$-resonance with respect to type~(a) leads to a suppression factor of $\Gamma_\tau/m_\tau \sim 10^{-12}$ in the contribution to the integrated cross section.
This large suppression cannot be compensated by new gauge-boson resonances, which imply enhancement factors of $M_{V}/\Gamma_V \sim 40$ ($V = \mathrm{W},\mathrm{Z}$) for each W/Z resonance. 
Note also that this enhancement from W/Z-boson resonances is largely mitigated by the fact that higher photon energies are required for their production.
As shown in Fig.~4 of Ref.~\cite{Shao:2022cly}, the probability to produce photons with a centre-of-mass energy $\sqrt{s_{\gamma\gamma}}$ decreases rapidly with $\sqrt{s_{\gamma\gamma}}$.
It is thus clear that the by far dominating production channel for this final state at typical LHC energies is the production of a resonant $\tau$-lepton pair.

Precision calculations for resonance processes including radiative corrections 
are complicated by the fact that the resonance pattern in propagators emerges
from a Dyson summation of self-energy corrections, {\it i.e.} particular care has to be taken
w.r.t.\ questions of gauge invariance and the cancellation of divergences
(see, {\it e.g.}, Ref.~\cite{Denner:2019vbn} and references therein).
However, owing to the large suppression of the contributions with less than two $\tau$-resonances in our
considered process, we can safely neglect any cross-section contribution that does not feature
two $\tau$-resonances and
reduce the matrix-element evaluation to the contributions from diagrams of type~(a).
The so-called {\it leading-pole approximation} provides the natural framework to implement this idea,
since it consistently identifies all resonance enhanced contributions to matrix elements and cross sections, 
and performs an expansion around the resonance poles. 
It allows for a consistent inclusion of radiative corrections in the vicinity of the resonance(s),
{\it i.e.} for invariant masses within a window of some decay widths at the resonating particle's
mass.
Owing to the smallness of the ratio between the decay width of the $\tau$-lepton and its mass,  
$\Gamma_\tau/m_\tau \sim 10^{-12}$, in our case 
the resonance pole expansion can be further simplified to the NWA without loss of precision.
In the NWA,
the square of the denominator of the $\tau$-leptons propagators is approximated by $\delta$-functions via the identity
\begin{align} \label{eq:id_propagator}
  \frac{1}{|p^2 - m_{\tau}^2 + \ri m_{\tau}\Gamma_{\tau}|^2} 
  \,\, \asymp{\Gamma_\tau \ll m_\tau} \,\,
  \frac{\pi}{m_{\tau}\Gamma_{\tau}}\delta(p^2 - m_{\tau}^2),
\end{align}
where $p$ is the momentum carried by the propagator.
Thus, in NWA the polarization-summed squared amplitude is evaluated according to
\begin{align} \label{eq:ME2}
  \overline{| \M |^2} 
  \,\, \underset{\mathrm{NWA}}{\longrightarrow} \,\,
  \bigg(\frac{\pi}{m_{\tau}\Gamma_{\tau}}\bigg)^2 \;
  \overline{|\widetilde{\M}|^2} \, \delta(p_{\tau}^2 - m_{\tau}^2)\delta(\bar{p}_{\tau}^2 - m_{\tau}^2),
\end{align}
where the momenta $p_{\tau}$ and $\bar{p}_{\tau}$ correspond to the momenta of the resonant $\tau^-$- and $\tau^+$-leptons, respectively,
and $\widetilde{\M}$ is the product of the subamplitudes of the $\tau$-pair production and $\tau$-decay subprocesses. 
Note that, in this approximation, the squared amplitude gives a non-zero contribution only in the phase-space regions where the $p^2_\tau = \bar{p}^2_\tau = m^2_\tau$ and, thus, the amplitude $\widetilde{\M}$ is evaluated for on-shell $\tau$-leptons.
The NWA divides the process into the on-shell production of a $\tau$-lepton pair and their decays.
Some details of using the NWA on the phase-space integrations in NLO predictions
are discussed in App.~\ref{app:NWA_PS}.


In the {\it naive} version of the NWA, the polarization-summed
square of the amplitude $\widetilde{\M}$ is evaluated as
\begin{align} \label{eq:ME2_NWA}
  \overline{|\widetilde{\M}_{\text{NWA}}|^2} = 
  \overline{|\M_{\mathrm{P}}|^2} \;
  \overline{|\M_{\mathrm{D}}|^2} \;
  \overline{|\M_{\overline{\mathrm{D}}}|^2},
\end{align}
where the subscripts P, D, and $\overline{\text{D}}$, refer to on-shell $\tau$-pair production, leptonic $\tau^-$-decay, and leptonic $\tau^+$-decay, respectively. 
The overlines indicate that the squared amplitudes for the individual subprocesses are considered for unpolarized external states.
In particular, this means that if the naive NWA is employed, the spin information of the produced $\tau$-leptons is not transferred to their decays, \ie spin correlations between the produced $\tau$-leptons are neglected.

These spin-correlation effects can be included in the calculation using an improved version of the NWA, for which the amplitude $\widetilde{\M}$ is evaluated as
\begin{align} \label{eq:ME2_iNWA}
  \widetilde{\M}_{\text{iNWA}} = \sum_{\sigma,\bar{\sigma}}\M_{\text{P},\sigma\bar{\sigma}}\M_{\text{D},\sigma}\M_{\overline{\text{D}},\bar{\sigma}},
\end{align}
and squared afterwards.
The sum $\sum_{\sigma,\bar{\sigma}}$ runs over all possible values for the helicities $\sigma$, $\bar{\sigma}$ of the $\tau^-$- and $\tau^+$-leptons, respectively.
Note that the dependence of the matrix elements on the polarizations of other external states has been suppressed to simplify the notation. 
For the evaluation of the matrix elements using the improved NWA, it is essential to choose the same 
bases for wave functions (\ie the same set of Dirac spinors, including identical phase conventions)
for the $\tau$-leptons required in the production and in the decays, and all matrix elements have to be evaluated in the same reference frame. 
This is particularly important to respect the Lorentz invariance of the matrix elements.
We consistently evaluate the matrix elements according to the conventions of Ref.~\cite{Dittmaier:1998nn}.

\subsection{Next-to-leading-order corrections}
\label{sec:NLO}

\subsubsection*{Naive and improved narrow-width approximation}

At LO, the use of the naive or improved NWA to evaluate the squared amplitude $|\widetilde{\M}|^2$ consists in using the LO amplitudes for each of the subprocesses 
according to Eq.~(\ref{eq:ME2_NWA}) or (\ref{eq:ME2_iNWA}), respectively, 
\ie $\M_{X} = \M^{(0)}_{X}$, where $X=\mathrm{P},\mathrm{D},\overline{\mathrm{D}}$, with the superscript $(0)$ indicating LO amplitudes.
At NLO, we first identify all doubly resonant contributions.
Note that not all NLO corrections to $\processmue$ can be factorized into independent parts for $\tau$-pair production and subsequent decays. 
We have to distinguish between the so-called {\it factorizable} and {\it non-factorizable} contributions.
The corrections are called {\it factorizable} if they comprise corrections to the production or to the decay subprocesses with on-shell kinematics for the resonant $\tau$-leptons. 
The remaining resonant corrections form the {\it non-factorizable} contributions.

Different strategies and important details of the decomposition of one-loop and real-emission amplitudes into factorizable and non-factorizable contributions can be found in 
Refs.~\cite{Beenakker:1997ir,Denner:1997ia,Dittmaier:2015bfe}. 
Generic results on virtual non-factorizable corrections are given, \emph{e.g.}, in Ref.~\cite{Dittmaier:2015bfe}.
If both, virtual and real corrections are treated within the leading-pole approximation~\cite{Denner:2019vbn,Stuart:1991xk,Aeppli:1993rs}, the sum of virtual and real non-factorizable corrections is expected to be rather small, as has been confirmed in explicit applications to single~\cite{Dittmaier:2014qza} and double resonances~\cite{Beenakker:1997ir,Denner:1997ia}. 
This suppression is due to the fact that non-factorizable corrections are only resonant for soft photon
exchange or emission, \ie they originate from photons with energies of the order of the
width of the resonance and can only be sizable within an invariant-mass window of this size around the resonance.
In observables where the invariant masses of the resonances are integrated over, the full non-factorizable corrections even reduce the non-resonant contributions, as pointed out already in Ref.~\cite{Fadin:1993kt}.
Since the NWA only considers factorizable contributions, the smallness of non-factorizable contributions in comparison with the factorizable contributions is a basic requirement for the use of the (naive or improved) NWA in the calculation of NLO EW corrections to $\processmue$.
In this section, we exemplify the classification into factorizable and non-factorizable corrections for the NLO EW corrections to $\processmue$, which are of $\order{\alpha^7}$. 

\begin{figure}
    \centering{
    \raisebox{0pt}{\includegraphics[width=0.32\textwidth]{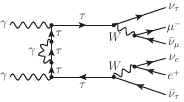}}
    \raisebox{0pt}{\includegraphics[width=0.32\textwidth]{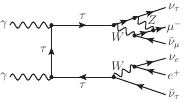}}
    \raisebox{0pt}{\includegraphics[width=0.32\textwidth]{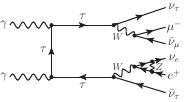}}
    \newline
    (a) \hspace{4.4cm} (b) \hspace{4.5cm}  (c) \hspace{-2cm}
    \newline   
    \raisebox{0pt}{\includegraphics[width=0.32\textwidth]{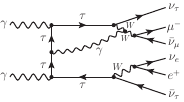}}
    $\qquad$ 
    \raisebox{0pt}{\includegraphics[width=0.32\textwidth]{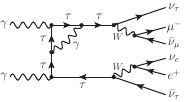}}
    \newline
    (d) \hspace{5.5cm} (e) \hspace{1.8cm}
    }
    \caption{Examples of one-loop Feynman diagrams for the process $\processmue$. 
    Diagrams (a), (b), and (c) belong to the manifestly factorizable contributions. 
    Diagram (d) is a manifestly non-factorizable contribution. 
    Diagram (e) contains a factorizable and a non-factorizable contribution.}
    \label{fig:NLOV}
\end{figure}

In Fig.~\ref{fig:NLOV}, examples of one-loop diagrams are given. 
Diagrams~(a), (b), and (c) are examples for diagrams that can be directly split into 
parts for the subprocesses of $\tau$-pair production and the subsequent decays and, therefore, they are part of the factorizable contributions.
More technically, they are {\it manifestly factorizable} contributions, meaning that the loop can be attributed to
exactly one of the subprocesses.
On the other hand, diagram (d) is an example of a one-loop correction for which the separation between production and decays obviously cannot be done, \ie it is a {\it manifestly non-factorizable} contribution.
Note that this classification is not always that straightforward.  
There are some diagrams for which the loop involves at least one coupling with the resonant $\tau$-leptons, see \eg diagram (e).
For such diagrams, partial fractioning can be used to the split them into a factorizable and a non-factorizable contribution.
For instance, the following identity can be used in (e) to separate it into a term leading to a factorizable contribution and a term leading to a non-factorizable contribution,
\begin{align} \label{eq:partial_fractioning}
  \frac{1}{[(p_\tau+k)^2 - m_\tau^2][p_\tau^2 - m_\tau^2]} =
  \frac{1}{(2p_\tau k)[p_\tau^2 - m_\tau^2]} - 
  \frac{1}{[(p_\tau+k)^2 - m_\tau^2](2p_\tau k)} ,
\end{align}
where $k$ and $p_\tau$ denote the momentum of the photon and of the $\tau^-$-lepton after photon emission, respectively.
Note that the various propagator factors on the r.h.s.\ of Eq.~(\ref{eq:partial_fractioning}) determine
which momentum ($p_\tau$ or $p_\tau+k$) is set on-shell in the subamplitudes for production or
decay. To be concrete, the first term on the r.h.s.\ of Eq.~(\ref{eq:partial_fractioning}) 
corresponds to radiation during the production, \ie the production subamplitude is evaluated for
$p_\tau^2 = m_\tau^2$ and the factor $1/(2p_\tau k)$ is the remnant of the first propagator
factor on the l.h.s.\ for $p_\tau^2 = m_\tau^2$. Likewise, the second term on the r.h.s.\ 
describes radiation during the decay subprocess, and the decay subamplitude is evaluated for
$(p_\tau+k)^2 = m_\tau^2$, with $-1/(2p_\tau k)$ representing the remnant of the second
propagator on the l.h.s. 
Applying this decomposition to diagram~(e) of Fig.~\ref{fig:NLOV}, in which the other end of the photon
propagator is linked to the production subprocess, the first contribution to the r.h.s.\ of
Eq.~(\ref{eq:partial_fractioning}) leads to a factorizable correction to the production, while the
second term leads to a non-factorizable contribution linking the resonance to the production.

As previously mentioned, only factorizable contributions are considered within the (naive or improved) NWA. 
Thus, the NLO EW correction from one-loop amplitudes to the polarization-summed 
square of $\widetilde{\M}$ can be split as
\begin{align} \label{eq:ME2_1L}
  \overline{\big|\delta\widetilde{\M}^{(1)}\big|^2} 
  = \overline{\big|\delta\widetilde{\M}^{(1,\mathrm{P})}\big|^2} \,
  + \, \overline{\big|\delta\widetilde{\M}^{(1,\mathrm{D})}\big|^2} \,
  + \, \overline{\big|\delta\widetilde{\M}^{(1,\overline{\mathrm{D}})}\big|^2},
\end{align}
where the superscript $(1)$ indicates that they correspond to one-loop corrections. 
The superscripts $\mathrm{P}$, $\mathrm{D}$, and $\overline{\mathrm{D}}$, denote if the loop belongs to $\tau$-pair production, to the $\tau^-$-lepton decay, or to the $\tau^+$-lepton decay.
While in the naive NWA the one-loop corrections are evaluated as
\begin{align} \label{eq:ME2_1L_NWA}
  \overline{\big|\delta\widetilde{\M}^{(1,\mathrm{P})}_\mathrm{NWA}\big|^2}  
  &= 2\Re\big\{\overline{\M^{(0)^*}_{\mathrm{P}}\M^{(1)}_{\mathrm{P}}}\big\} \;
     \overline{\big|\M^{(0)}_{\mathrm{D}}\big|^2}  \;
     \overline{\big|\M^{(0)}_{\overline{\mathrm{D}}}\big|^2}, 
  \notag \\
  \vh 
  \overline{\big|\delta\widetilde{\M}^{(1,\mathrm{D})}_\mathrm{NWA}\big|^2}  
  &= \overline{\big|\M^{(0)}_{\mathrm{P}}\big|^2}  \;
     2\Re\big\{\overline{\M^{(0)^*}_{\mathrm{D}}\M^{(1)}_{\mathrm{D}}}\big\} \;
     \overline{\big|\M^{(0)}_{\overline{\mathrm{D}}}\big|^2}, 
  \notag \\
  \overline{\big|\delta\widetilde{\M}^{(1,\overline{\mathrm{D}})}_\mathrm{NWA}\big|^2}  
  &= \overline{\big|\M^{(0)}_{\mathrm{P}}\big|^2}  \;
     \overline{\big|\M^{(0)}_{\mathrm{D}}\big|^2}  \;
     2\Re\big\{\overline{\M^{(0)^*}_{\overline{\mathrm{D}}}
               \M^{(1)}_{\overline{\mathrm{D}}}}\big\},
\end{align}
in the improved version of the NWA they are given by
\begin{align} \label{eq:ME2_1L_iNWA}
  \overline{\big|\delta\widetilde{\M}^{(1,X)}_\mathrm{iNWA}\big|^2}   
  = 2\Re\big\{\overline{\widetilde{\M}^{(0)^*}_{\mathrm{iNWA}}
    \widetilde{\M}^{(1,X)}_{\mathrm{iNWA}}}\big\},
\end{align}
with
\begin{align} \label{eq:ME_1L_iNWA}
\widetilde{\M}_{\text{iNWA}}^{(0)} 
  &= \sum_{\sigma,\bar{\sigma}}
      \M^{(0)}_{\text{P},\sigma\bar{\sigma}}
      \M^{(0)}_{\text{D},\sigma}
      \M^{(0)}_{\overline{\text{D}},\bar{\sigma}},
  \notag \\
  \widetilde{\M}_{\text{iNWA}}^{(1,\text{P})} 
  &= \sum_{\sigma,\bar{\sigma}}
      \M^{(1)}_{\text{P},\sigma\bar{\sigma}}
      \M^{(0)}_{\text{D},\sigma}
      \M^{(0)}_{\overline{\text{D}},\bar{\sigma}},
  \notag \\
  \widetilde{\M}_{\text{iNWA}}^{(1,\text{D})} 
  &= \sum_{\sigma,\bar{\sigma}}
     \M^{(0)}_{\text{P},\sigma\bar{\sigma}}
     \M^{(1)}_{\text{D},\sigma}
     \M^{(0)}_{\overline{\text{D}},\bar{\sigma}}, 
  \notag \\
  \widetilde{\M}_{\text{iNWA}}^{(1,\overline{\text{D}})} 
  &= \sum_{\sigma,\bar{\sigma}}
     \M^{(0)}_{\text{P},\sigma\bar{\sigma}}
     \M^{(0)}_{\text{D},\sigma}
     \M^{(1)}_{\overline{\text{D}},\bar{\sigma}}.
\end{align}

\begin{figure}
    \centering{
    \raisebox{0pt}{\includegraphics[width=0.49\columnwidth]{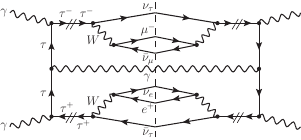}}
    \hspace{-0.1cm}
    \raisebox{0pt}{\includegraphics[width=0.49\columnwidth]{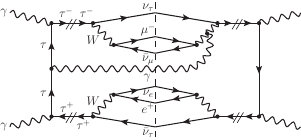}}
    \newline
    \wm(a) \hspace{7.25cm} (b) 
    }
    \caption{
    Examples of real-emission interference diagrams leading to (a) a factorizable correction to $\tau$-pair production and (b) a non-factorizable correction.
    The double line indicates the resonance propagator, \ie
    the momenta of these lines are set on-shell in the rest of the diagrams. The dashed line indicates a unitarity cut, \ie the whole diagram is the product of the subdiagram on the left and the complex conjugate of the subdiagram on the right.}
    \label{fig:RE}
\end{figure}

Similarly to the virtual one-loop corrections, real-emission corrections are classified into factorizable and non-factorizable contributions. 
The factorizable contributions to the real corrections
comprise the sum of all diagrams where the photon is emitted from the same subprocess.
Non-factorizable contributions are given by the interference between diagrams where the photon is emitted from different subprocesses.
As for the one-loop amplitude, particular care is needed when the photon is radiated off one of the resonant $\tau$-leptons.
In this case, partial fractioning~(\ref{eq:partial_fractioning}) is again employed to split such a diagram,
doubling the number of contributions for each application of partial fractioning.
Figure~\ref{fig:RE} illustrates two interference contributions to the squared real emission amplitude:
Diagram~(a) corresponds to emission during $\tau$-pair production and
diagram~(b) to a manifestly non-factorizable contribution.

Within the NWA, the factorizable real-emission corrections to the square of $\widetilde{\M}$ can be divided as
\begin{align} \label{eq:ME2_RE}
  \overline{\big|\widetilde{\M}^{(\gamma)}\big|^2} 
  = \overline{\big|\widetilde{\M}^{(\gamma,\mathrm{P})}\big|^2} \,
  + \, \overline{\big|\widetilde{\M}^{(\gamma,\mathrm{D})}\big|^2} \,
  + \, \overline{\big|\widetilde{\M}^{(\gamma,\overline{\mathrm{D}})}\big|^2},
\end{align}
where the superscript $(\gamma)$ indicates that they are real-emission amplitudes and the superscripts $\mathrm{P}$, $\mathrm{D}$, and $\overline{\mathrm{D}}$, indicate if the emission occurs in the on-shell production of the $\tau$-pair, in the $\tau^-$-lepton decay, or in the $\tau^+$-lepton decay.
If spin-correlation effects are neglected, \ie using the naive NWA, the contributions in the previous equation are given by
\begin{align} \label{eq:ME2_RE_NWA}
  \overline{\big|\widetilde{\M}^{(\gamma,\mathrm{P})}_\mathrm{NWA}\big|^2}  
  &= \overline{\big|\M^{(\gamma)}_{\mathrm{P}}\big|^2}\;
     \overline{\big|\M^{(0)}_{\mathrm{D}}\big|^2} \;
     \overline{\big|\M^{(0)}_{\overline{\mathrm{D}}}\big|^2}, 
  \notag \\
  \vh \overline{\big|\widetilde{\M}^{(\gamma,\mathrm{D})}_\mathrm{NWA}\big|^2}  
  &= \overline{\big|\M^{(0)}_{\mathrm{P}}\big|^2} \;
     \overline{\big|\M^{(\gamma)}_{\mathrm{D}}\big|^2} \;
     \overline{\big|\M^{(0)}_{\overline{\mathrm{D}}}\big|^2}, 
  \notag \\
  \overline{\big|\widetilde{\M}^{(\gamma,\overline{\mathrm{D}})}_\mathrm{NWA}\big|^2}  
  &= \overline{\big|\M^{(0)}_{\mathrm{P}}\big|^2}  \;
     \overline{\big|\M^{(0)}_{\mathrm{D}}\big|^2} \;
     \overline{\big|\M^{(\gamma)}_{\overline{\mathrm{D}}}\big|^2}.
\end{align} 
On the other hand, when spin correlations between the $\tau$-leptons are included, \ie employing the improved version of the NWA, the amplitudes in the right-hand side of Eq.~(\ref{eq:ME2_RE}) are evaluated according to
\begin{align} \label{eq:ME_RE_iNWA}
  \widetilde{\M}_{\text{iNWA}}^{(\gamma,\text{P})} 
  &= \sum_{\sigma,\bar{\sigma}}
     \M^{(\gamma)}_{\text{P},\sigma\bar{\sigma}}
     \M^{(0)}_{\text{D},\sigma}
     \M^{(0)}_{\overline{\text{D}},\bar{\sigma}}, 
  \notag \\
  \widetilde{\M}_{\text{iNWA}}^{(\gamma,\text{D})} 
  &= \sum_{\sigma,\bar{\sigma}}
     \M^{(0)}_{\text{P},\sigma\bar{\sigma}}
     \M^{(\gamma)}_{\text{D},\sigma}
     \M^{(0)}_{\overline{\text{D}},\bar{\sigma}}, 
  \notag \\
  \widetilde{\M}_{\text{iNWA}}^{(\gamma,\overline{\text{D}})} 
  &= \sum_{\sigma,\bar{\sigma}}
     \M^{(0)}_{\text{P},\sigma\bar{\sigma}}
     \M^{(0)}_{\text{D},\sigma}
     \M^{(\gamma)}_{\overline{\text{D}},\bar{\sigma}},
\end{align}
and squared afterwards.
%
%
The fact that for each subprocess, real-emission amplitudes are evaluated with on-shell momenta for the resonant $\tau$-leptons, leads to a different infrared (IR) structure of the amplitude as compared with the one of the original off-shell matrix element.%
\footnote{In fact the non-commutativity of the IR-limit of photon exchange or emission and the limit of going
on resonance is the reason why non-factorizable corrections can lead to resonant contributions.}
This fact has to be addressed in the construction of a proper IR subtraction term, which is described in detail in App.~\ref{app:NWA_DS}.


\subsubsection*{Classification of NLO corrections}

To provide a detailed study of the NLO EW corrections to $\processmue$, we split the total NLO EW correction into gauge-invariant subsets.
A first distinction is made between corrections to $\tau$-pair production, $\Delta\sigma^{\NLO}_{\text{P}}$, to $\tau^-$-lepton decay, $\Delta\sigma^{\NLO}_{\text{D}}$, and to $\tau^+$-lepton decay, $\Delta\sigma^{\NLO}_{\overline{\text{D}}}$.

Each of these contributions can be further split into gauge-invariant subcontributions. 
Since the SM could be formulated for any number of fermion generations, the sum of all closed fermion-loop corrections provides a gauge-invariant contribution for each fermion generation as long as the generation does not mix with the others. 
Accordingly, the corrections to the subprocesses are divided into corrections including closed fermion loops, $\Delta\sigma^{\NLO}_{\text{ferm}}$, and the rest, to which we refer to as bosonic corrections, $\Delta\sigma^{\NLO}_{\text{bos}}$.

Furthermore, since there are no weak gauge bosons in the LO diagrams for $\tau$-pair production, the bosonic corrections can be further split into QED, $\Delta\sigma^{\NLO}_{\text{QED}}$, and genuinely weak corrections, $\Delta\sigma^{\NLO}_{\text{weak}}$. 
The QED corrections include the exchange of virtual photons between $\tau$-leptons in all loop diagrams and renormalization constants, as well as the real emission of a photon. 
The weak corrections account for the remaining bosonic corrections. 
In the literature, the closed fermion loops are often included in the weak corrections. 
In this work, they are considered separately to emphasize the specific contribution of closed fermion loops to the NLO corrections, which depends on the choice of input-parameter scheme for the electromagnetic coupling.

\begin{figure}
    \centering{
    \raisebox{0pt}{\includegraphics[width=0.3\columnwidth]{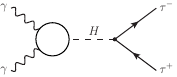}}
    \caption{
    One-loop corrections to $\gamma\gamma\to\tau^+\tau^-$ with a resonant
    Higgs-boson propagator. 
    The blob stands for all one-loop subdiagrams contributing to the $\gamma\gamma H$ vertex.}
    \label{fig:aa_H_tautau}
    }
\end{figure}

The corrections to $\tau$-pair production receive contributions with a potentially resonant 
Higgs-boson propagator, see Fig.~\ref{fig:aa_H_tautau}.
A proper treatment of these corrections requires an isolation of this one-loop-induced
resonant contribution to the amplitude, the introduction of a finite Higgs width therein,
and the inclusion of its square in the squared amplitude, 
see \eg Refs.~\cite{Denner:1995jv,Denner:1995ar}. 
In our case, this resonant contribution is highly suppressed by the drop of the
photon flux for high momentum transfer, so that the whole contribution does not play any role phenomenologically.
To estimate this suppression, we compare the LO cross section for 
$\gamma\gamma\to\tau^+\tau^-$ (see Table~\ref{tbl:tautau_prd} below)
with the LO cross section $\sigma^\LO_{\mathrm{H}\to\tau\tau}$ 
for the Higgs-mediated process $\gamma\gamma\to H\to\tau^+\tau^-$,
\begin{align}
  \sigma^\LO_{\mathrm{H}\to\tau\tau} \approx 20 \,\mathrm{fb},
\end{align}
which is suppressed by more than ten orders of magnitude compared to the result given in Table~\ref{tbl:tautau_prd}.
Assuming an energy resolution of $\sim 1\GeV$, 
the Higgs resonance at $m_{\tau\tau} = 125\GeV$
in the distribution of the invariant mass $m_{\tau\tau}$ of the $\tau$-pair
is still suppressed by three orders of magnitude w.r.t.\ the continuum and, thus, phenomenologically
negligible.

The NLO EW corrections to the decays of the $\tau$-leptons can be split into fermionic and bosonic parts in the same way. 
However, the $\tau$-lepton decay is a charged-current process, \ie induced W-boson exchange at LO, and the corresponding bosonic corrections cannot be split into QED and weak corrections in a gauge-invariant way.

\subsection{Technical aspects of the NLO calculation and employed tools}
\label{sec:tools}


For the presented calculation, a Monte Carlo (MC) program dubbed 
\MonteTau has been developed based on the tools and concepts described in the following. 
\MonteTau is a \cpp program which provides NLO EW predictions for the process $\process$ in UPCs of heavy ions. 
Moreover, it optionally provides NLO EW predictions for the subprocesses $\gamma\gamma\to L^+L^-$ and $L^- \to \ell^- \nu_L \bar{\nu}_{\ell}$ individually, with $L = \tau, \mu, e$ and $\ell = \mu, e$ (if the combination is allowed).
%
%
For photon--photon scattering, the program offers the possibility to generate initial-state photons 
radiated off heavy ions, \eg lead ions, or to take monochromatic initial-state photons.
The photon flux factor, which parametrizes the radiation coming from the heavy ions, is calculated using the library \gammaUPC~\cite{Shao:2022cly}.


As explained above, the naive or improved NWA is used to calculate the squared matrix elements of the hard process.
At tree level, the matrix elements have been calculated for $\gamma\gamma\to L^+L^-$ and for $L^- \to \ell^- \nu_L \bar{\nu}_{\ell}$ using the spinor-helicity formalism of Ref.~\cite{Dittmaier:1998nn}. 
This formalism has also been used to obtain analytical expressions for the matrix elements of the real-emission corrections to $L^- \to \ell^- \nu_L \bar{\nu}_{\ell}$. 
The other matrix elements have been generated with the help of the \Mathematica~\cite{Mathematica} packages \FeynArts~\cite{Kublbeck:1990xc, Hahn:2000kx} and \FormCalc~\cite{Hahn:1998yk}. 
Note that \FormCalc\ follows the definition of Ref.~\cite{Dittmaier:1998nn} for 
Dirac spinors for helicity states as well.

For the evaluation of the loop integrals the \collier library~\cite{Denner:2014gla,Denner:2016kdg} is used.
%
%
The integration over the phase space of the final-state particles and over the energies of the 
initial-state photons is performed using the \VEGAS algorithm~\cite{Lepage:1977sw, Lepage:1980dq}
employing its implementation of the \gsl~\cite{Galassi:2019czg}.


\MonteTau makes use of an on-shell renormalization scheme to renormalize the parameters and fields of the EW part of the SM (see Refs.~\cite{Denner:2019vbn,Denner:1991kt}) and handles the IR singularities with the dipole subtraction formalism~\cite{Dittmaier:1999mb,Dittmaier:2008md,Basso:2015gca}.
Since on-shell renormalization conditions are used for all the parameters appearing in the calculation, there is no dependence on the renormalization scale in the presented results. 
Furthermore, there are no collinear singularities arising from the radiation off the heavy ions 
which remain intact in UPCs, and the employed photon spectra do not depend on any arbitrary factorization
scale.
Thus, there is neither a dependence on the renormalization scale nor on a factorization scale in the results.


\MonteTau offers the possibility to chose between either the $\alpha(0)$-scheme 
and the \GF-scheme as the input-parameter scheme for the electromagnetic coupling in each subprocess independently.
In particular, this flexibility allows for the use of a mixed input-parameter scheme in predictions for the full process as in our best prediction presented below.
Moreover, \MonteTau includes the possibility to renormalize the weak-gauge-boson masses and fields within the complex-mass scheme~\cite{Denner:2019vbn,Denner:2005fg}. 
By default real pole masses are used.

\subsubsection*{Checks on the calculation}

To ensure the correctness of our calculation, we have carried out several consistency checks:
\begin{itemize}
  \item All the derived analytical expressions for scattering amplitudes have been numerically validated against matrix elements generated using \FeynArts and \FormCalc.
  \item All squared matrix elements for unpolarized final states have been numerically verified to be invariant under Lorentz transformations. 
  \item All one-loop matrix elements have been checked to be ultraviolet (UV) finite after renormalization.
  \item The ratio between the squared real-emission matrix element and the dipole subtraction function has been observed to approach unity in the IR limits for the entire process and for the individual subprocesses.
  \item Results for $\tau$-pair production for fixed centre-of-mass-energy and for the 
$\tau$-lepton decay have been compared at NLO EW with an independent computation performed with \McMule~\cite{Banerjee:2020rww}.
  \item Results for on-shell $\tau$-pair production in UPCs have been compared at NLO EW for massless $\tau$-leptons with an independent calculation performed with {\sc MoCaNLO} using the matrix-element generator \Recola~\cite{Denner:2017wsf}.
  \item The results for the $2\to 6$ process have been compared with results from an independent calculation at LO for massless final-state particles in the NWA neglecting the spin-correlation effects using the matrix-element generator \Recola~\cite{Denner:2017wsf}.
  The in-house Monte Carlo used for this check has already been used in various contexts~\cite{Gavin:2014yga,Cavasonza:2014xra,Cavasonza:2016qem,Agliardi:2022ghn,Denner:2023grl}.
  \item The predictions for the improved version of the NWA have been compared at 
LO with predictions based on an off-shell matrix element that keeps only the diagrams 
involving $\tau$-pair production (evaluated with on-shell kinematics).
\end{itemize}

\subsection{Set-up of the calculation}
\label{sec:setup}

\subsubsection*{Input-parameter scheme}

In EW precision calculations, it is crucial to use an appropriate input-parameter scheme in order to absorb universal corrections into LO predictions as much as possible and to avoid the appearance of ill-defined quantities (such as masses of the light quarks) in the corrections. 

The $\alpha(0)$-scheme~\cite{Denner:2019vbn} is preferred for electromagnetic
processes without internal gauge-boson lines in the LO diagrams, such as 
$\gamma\gamma\to\tau^+\tau^-$.
At NLO the combination $\delta Z_{AA} + 2\delta Z_{e}$ of the wave-function renormalization constant $\delta Z_{AA}$ of the photon and the charge renormalization constant $\delta Z_e$ occurs, \ie exactly the combination that vanishes in pure QED for on-shell renormalization of the electric charge in the Thomson limit.
In other words, these processes do not receive any universal correction from coupling renormalization, reflecting the fact that on-shell photons effectively couple with $\alpha(0)$.

The use of the $\alpha(0)$-scheme for processes featuring W bosons and no external photons, 
such as $\tau^-\to e^-\nu_{\tau}\bar{\nu}_{e}$, leads to large logarithmic corrections arising from the running of $\alpha$ from momentum transfer $0$ up to the electroweak scale. 
These corrections contain mass-singular terms of the form $\alpha\ln{m_f}$ from each light fermion $f$. 
Moreover, the leading corrections to the $\rho$-parameter, which is dominated by top-quark loops, are also present at NLO arising from the on-shell renormalization of the weak mixing angle $\theta_{\mathrm{w}}$. 
On the other hand, the \GF-scheme~\cite{Denner:2019vbn}, which defines the EW coupling via the LO relation
\begin{equation} \label{eq:GF_scheme}
  \alpha_{\GF} = \frac{\sqrt{2}}{\pi} G_\mu \MW^2 \left( 1 - \frac{\MW^2}{\MZ^2} \right),
\end{equation}
absorbs the universal corrections to $\alpha(M_\mathrm{Z}^2)/\sin^2{\theta_{\mathrm{w}}}$ into the value of $\alpha$, is preferred for these processes. 
Note that if the \GF-scheme was used for processes featuring external photons, the running of $\alpha$ and the corrections to the $\rho$-parameter would be erroneously absorbed into $\alpha$, so that the relative corrections would receive the large shift to compensate these misplaced corrections.

In order to illustrate the importance of the choice of input-parameter scheme for the electromagnetic coupling, we consider on-shell $\tau$-pair production and on-shell leptonic $\tau$-decay separately.
In Table~\ref{tbl:tautau_prd}, we show the results for the inclusive cross section for on-shell $\tau$-pair production in UPCs without restrictions on the phase space, \ie we use a setup close to the one of Ref.~\cite{Jiang:2024dhf}. 
The results are given at LO and NLO accuracy for the $\alpha(0)$- and \GF-schemes.
\begin{table}
  \centering{
    \begin{tabular}{c|c|c|c|c}
     \hl $\gamma\gamma\to\tau^+\tau^-$ 
& \multicolumn{2}{c|}{$\alpha(0)$-scheme} & \multicolumn{2}{c}{\GF-scheme}  \\ 
     \hl
& $\sigma$ or $\Delta \sigma$ $[\text{mb}]$ & $\delta \, [\%]$ &  $\sigma$ or $\Delta \sigma$ $[\text{mb}]$ & $\delta \, [\%]$ \\ 
     \hline \hline
     $\sigma^{\LO}                       $ \hl & \wm$1.063(2)$         & -                     & \wm$1.136(3)$ & -            \\ \hline \hline
     $\Delta \sigma^{\NLO}_{\text{QED}}  $ \hl & \wm$0.010(3)$         & $0.94(3)$             & \wm$0.012(1)$ & \wm$1.08(6)$ \\ \hline 
     $\Delta \sigma^{\NLO}_{\text{weak}} $ \hl & $9.1(7)\times10^{-8}$ & $8.5(6)\times10^{-6}$ & $-0.009(3)  $ & $-0.84(1)$   \\ \hline
     $\Delta \sigma^{\NLO}_{\text{ferm}} $ \hl & $6.6(1)\times10^{-7}$ & $6.2(6)\times10^{-5}$ & $-0.058(1)  $ & $-5.10(2)$   \\ \hline \hline
     $\sigma^{\NLO}                      $ \hl & \wm$1.073(2)$         & $0.94(3)$             & \wm$1.081(3)$ & $-4.86(6)$   \\ 
    \end{tabular}
  }
  \caption{
  NLO corrections to $\tau$-pair production induced by UPCs of two lead ions with $\sqrt{s_{\text{PbPb}}}=5.02 \TeV$. 
  The first line with results gives the inclusive LO cross section.
  The following lines provide the QED, weak, and fermionic corrections.
  Finally, the last line gives the inclusive NLO EW cross section.
  The second column-like block shows the predictions using the $\alpha(0)$-scheme for the electromagnetic coupling, and the third block refers to the results using the \GF-scheme.
  For each scheme the corrections are given as absolute correction $\Delta\sigma$ to the cross section $\sigma$ and as relative correction factor $\delta = \sigma/\sigma_\LO -1$.
  The numbers in parentheses are the integration errors of the last digits.
  }
  \label{tbl:tautau_prd}
\end{table}
Our result of $0.9\%$ for the relative EW correction in the $\alpha(0)$-scheme
is consistent with the respective result of $1.0\%$
reported in Ref.~\cite{Shao:2025bma} for the somewhat different
Pb--Pb scattering energy of $\sqrt{s_{\text{PbPb}}}=5.52\TeV$.
On the other hand, our result of $-4.9\%$ on the relative EW correction in the \GF-scheme
show some difference to the corresponding result of $-3.1\%$ reported in
Ref.~\cite{Jiang:2024dhf}.\footnotemark
\footnotetext{Note that in Ref.~\cite{Jiang:2024dhf} the parametrization of the photon flux is different from the one used in the present work.
There, the survival probability for the heavy ions is assumed to be always one, while we do not make this assumption. Moreover, in Ref.~\cite{Jiang:2024dhf} the photon number density is parametrized using the EDFF approach~\cite{Cahn:1990jk} of the heavy ion, while for our results in Table~\ref{tbl:tautau_prd} the ChFF approach~\cite{Vidovic:1992ik} is used. 
Note also that the masses of the light quarks are not provided in Ref.~\cite{Jiang:2024dhf}, so that not all the input parameters could be matched.
}
While the weak and fermionic corrections are negligible in comparison to QED corrections in the calculation employing the $\alpha(0)$-scheme, they are of $\sim6\,\%$ in the computation using the \GF-scheme. 
As explained above, this large correction has its origin in misplaced corrections from the definition of the electromagnetic coupling and the universal corrections to $\sin{\theta_{\mathrm{w}}}$ originating from the $\rho$-parameter.
For completeness, we also include in Table~\ref{tbl:tau_dcy} results for the inclusive partial decay width of the $\tau$-lepton for the leptonic decay channel $\tau^-\to e^- \nu_\tau\bar{\nu}_e$.
\begin{table}
  \centering{
    \begin{tabular}{c|c|c|c|c}
     \hl $\tau^-\to e^- \nu_\tau\bar{\nu}_e$ 
& \multicolumn{2}{c|}{$\alpha(0)$-scheme} & \multicolumn{2}{c}{\GF-scheme}  \\ 
     \hl
& $\Gamma$ or $\Delta \Gamma$ $[\text{ns}^{-1}]$ & $\delta \, [\%]$ &  $\Gamma$ or $\Delta \Gamma$ $[\text{ns}^{-1}]$ & $\delta \, [\%]$         \\ 
     \hline \hline
     $\Gamma^{\LO}                       $ \hl & $573.35(8)$        & -                  & \wm$615.28(9)$        & -                        \\ \hline \hline
     $\Delta \Gamma^{\NLO}_{\text{bos}}  $ \hl & \wn\wn$2.18(1)$    & \,$0.38(1)$\,      & \wn\wn$-2.69(1)$      & $-0.44(1)$               \\ \hline
     $\Delta \Gamma^{\NLO}_{\text{ferm}} $ \hl & \wn$28.18(3)$      & \,$4.92(1)$\,      & $1.1(1)\times10^{-2}$ & $1.9(1)\times10^{-3}$    \\ \hline \hline
     $\Gamma^{\NLO}                      $ \hl & $603.71(9)$        & \,$5.30(1)$\,      & \wm$612.60(9)$        & $-0.44(1)$               \\ 
    \end{tabular}
  }
  \caption{
  NLO corrections to $\tau^-\to e^- \nu_\tau\bar{\nu}_e$ in the rest-frame of the $\tau$-lepton. 
  The first line with results gives the inclusive LO decay width.
  The following lines provide the bosonic and fermionic corrections.
  Finally, the last line gives the inclusive NLO EW decay width.
  The second column-like block shows the predictions using the $\alpha(0)$-scheme for the electromagnetic coupling, and the third block refers to the results using the \GF-scheme.
  For each scheme the corrections are given as absolute correction $\Delta\Gamma$ to the decay width $\Gamma$ and as relative correction factor $\delta = \Gamma/\Gamma_\LO -1$.
  The numbers in parentheses are the integration errors of the last digits.
  }
  \label{tbl:tau_dcy}
\end{table}
In this case, the optimal input-parameter scheme is the \GF-scheme,
while the $\alpha(0)$-scheme does not absorb the universal corrections
from the running of $\alpha$ and from corrections to the $\rho$-parameter into the
LO prediction.
This is mostly reflected in the closed fermion-loop corrections being four orders of magnitude smaller when using the \GF-scheme rather than the $\alpha(0)$-scheme. 
In view of the running effects in $\alpha$, 
the non-optimal input-parameter scheme just adapts the values of the light-quark masses to reproduce the experimental value for the hadronic vacuum polarization.
Of course, in an all-order calculation, the cross section and the partial decay width would be independent of the used input-parameter scheme (as long as the SM describes the muon decay correctly), so that the total NLO EW predictions do not differ significantly, in contrast to the LO predictions.

For processes with $n$ EW coupling factors $\alpha$ in the LO cross section that involve $l$ external photons and $(n-l)/2$ internal W bosons, \eg $\processmue$, the $n$ coupling factors $\alpha^n$ in the LO cross section can be consistently parametrized via a mixed scheme~\cite{Denner:2019vbn} to achieve the full absorption of the universal correction $\Delta\alpha$ describing the running of $\alpha$
and to absorb the leading corrections to the $\rho$-parameter into the EW couplings.
To this end, the $\alpha(0)$-scheme should be used for $l$ of the EW couplings and the \GF-scheme for the $n-l$ remaining EW couplings, so that the LO cross section scales like $\alpha(0)^l\alpha_{\GF}^{n-l}$.
For NLO EW calculations, the additional $\alpha$ factor appearing in the NLO EW correction can be chosen accordingly either as $\alpha(0)$ or as $\alpha_{\GF}$ since the difference between the two options is beyond NLO accuracy.

\subsubsection*{Numerical input}
\label{sec:inputs}

The results presented in the following are for UPCs of two lead ions ($Z = 82$) with a centre-of-mass energy of $\sqrt{s_{\text{PbPb}}} = 5.02 \TeV$.

The \EW\ coupling is fixed in the $\alpha(0)$-scheme to 
\begin{align}
  \alpha^{-1}(0) = 137.035999180, 
\end{align}
and in the \GF-scheme (see, \emph{e.g.}, Ref.~\cite{Denner:2019vbn}) upon 
%
%
the Eq.~(\ref{eq:GF_scheme}), 
with
\begin{equation}
  \GF    = 1.16638\times 10^{-5}\GeV^{-2}\;.
\end{equation}
In the \gammaUPC library, the used value for the \EW\ coupling is $\alpha^{-1} = 137.036$ because the initial-state photons are considered quasi-real, \emph{i.e.}\ the natural energy scale for the coupling is $q^2 = 0$.
For the hard-scattering process we use a mixed input-parameter scheme in which the $\alpha(0)$-scheme is employed for $\tau$-pair production and the \GF-scheme is used for $\tau$-lepton decays, consistently at LO and at NLO, \ie the factors of $\alpha$ in the relative EW corrections are chosen accordingly.

Moreover, the following masses and widths are used:
\begin{alignat}{2}
  \ME   &= 510.99895\keV,       &  \quad \quad \quad \MM &=  0.1056583755 \GeV,  \nonumber  \\
  \ML   &=   1.77686\GeV,       & \quad \quad \quad \Gamma_\tau &= 2.2673508677230445\times10^{-12}\GeV,  \nonumber \\
  \Mt   &=   173.0\GeV,          &  \quad \quad \quad \Mb   &=   4.92\GeV,      \nonumber \\
  \Mc   &=   1.51\GeV,          &  \quad \quad \quad \MH &=  125.0\GeV,  \nonumber \\
  \MZOS &=  91.1876\GeV,        & \quad \quad \quad \GZOS &= 2.4952\GeV,  \nonumber \\
  \MWOS &=  80.379\GeV,         & \GWOS &= 2.085\GeV . 
\end{alignat}
The values of the fermion masses are taken from Ref.~\cite{Beringer:1900zz}. 
Although the results do not depend on the values of the light quark masses, for the intermediate steps, the values $m_{\Pu,\Pd,\Ps} = 0.1 \GeV$ are used for the masses of the light quarks. 
These masses have been chosen such that the correct experimental value for $\alpha^{-1}(M_\mathrm{Z}^2) \approx 129$ is reproduced. 
The pole masses and widths of the W and Z bosons used in the numerical calculations are obtained from the measured on-shell (OS) values via~\cite{Denner:2019vbn,Bardin:1988xt}
\begin{equation}\label{eq:pole-mass-width}
  M_{\text{V}} = \frac{\MVOS}{\sqrt{1+(\GVOS/\MVOS)^2}}\;,\qquad
  \Gamma_{\text{V}} = \frac{\GVOS}{\sqrt{1+(\GVOS/\MVOS)^2}}\;,
\end{equation}
with ${\text{V}}=\PW, \PZ$.%
\footnote{The values of the widths $\Gamma_{\PZ/\PW}$ do not enter the results given in the following;
they were only relevant in the check performed within the complex-mass scheme.} 

We finally recall that the weak mixing angle $\theta_{\text{w}}$ involved in the EW coupling is fixed according to the all-order relation 
\begin{align}
  \text{c}_{\text{w}} \equiv \cos{\theta_{\text{w}}} = \frac{M_{\text{W}}}{M_{\text{Z}}},
\end{align}
because $M_{\text{W}}$ and $M_{\text{Z}}$ are used as independent input parameters.

\subsubsection*{Event selection}
\label{sec:selection}

The event selection mimics the one defined in Ref.\,\cite{ATLAS:2022ryk}, where the considered process was observed first with both $\tau$-leptons decaying leptonically. 
The experimental signature consists in two charged leptons of opposite sign and missing transverse energy. 
The cuts are
\begin{align}\label{eq:lepton}
\ptsub{\ell^\pm} >  4\GeV, \qquad \qquad |\eta_{\ell^\pm}| < 2.5,
\end{align}
where $p_{\rT}$ and $\eta$ are the transverse momentum and the pseudorapidity, respectively.
The transverse momentum and the pseudorapidity of a particle $i$ are, respectively, defined as
\begin{align}
  p_{\text{T},i} = \sqrt{p_{i,x}^2 + p_{i,y}^2}, 
  \qquad \qquad
  \eta_i = - \frac{1}{2} \ln\bigg[\tan\bigg(\frac{\theta_i}{2} \bigg)\bigg] ,
\end{align}
where $\theta_i$ is the polar angle of the particle $i$ in the laboratory frame. 
The $z$-axis is identified with the beam axis, and its positive direction is chosen according to the direction of flight of the positron, so that 
rapidity and pseudo-rapidity distributions are not necessarily symmetric under $\eta_i\to-\eta_i$. In particular,
the pseudorapidity of the positron is never negative in this set-up.


If not explicitly mentioned in the calculation of NLO EW corrections, collinear radiation off the positron and off the muon is treated inclusively via the following photon-recombination procedure: 
Final-state photons are recombined with the final-state charged leptons $\ell$ if
\begin{align}\label{eq:recombination}
  \Delta R_{\gamma\ell} < 0.1.
\end{align}
The angular distance $\Delta R_{ij}$ is defined as,
\begin{align} \label{eq:Delta_R}
  \Delta R_{ij} = \sqrt{(\Delta\phi_{ij})^2 + (\Delta\eta_{ij})^2},
\end{align}
where $\Delta\phi_{ij}$ and $\Delta\eta_{ij}$ are, respectively, the azimuthal and pseudorapidity differences between the $i$-th and $j$-th particles, 
\begin{align}
  \Delta\phi_{i j} 
              = \text{min}\{|\phi_i-\phi_j|, 2\pi-|\phi_i - \phi_j|\},
  \qquad
  \Delta\eta_{i j} =\eta_i - \eta_j.  
\end{align}
In the case where both final-state charged leptons fulfil condition~(\ref{eq:recombination}), the photon is recombined with the final-state charged lepton for which the associated value for $\Delta R_{\gamma\ell}$ is the smallest.
The recombined lepton--photon system is called \textit{dressed lepton} and carries a momentum
\begin{align}
  p_{\ell\gamma} = p_{\ell} + p_{\gamma},
\end{align}
where $p_{\ell\gamma}$, $p_\ell$, and $p_\gamma$ are the momenta carried by the dressed lepton, the final-state lepton, and the photon, respectively.

\section{Results}
\label{sec:results}

In this section, our results on the fiducial cross section for $\processmue$ induced by UPCs of lead ions as well as on the following differential distributions are given: 
the transverse momentum of the muon, $p_{\text{T},\mu}$, 
the absolute value of the pseudorapidity of the muon, $|\eta_{\mu}|$, 
the azimuthal angle distance between the muon and the positron, $\Delta \phi_{e\mu}$, 
and the pseudorapidity difference between the muon and the positron, $\Delta \eta_{e \mu}$.

First, we present our best SM prediction for the aforementioned observables, 
providing results at LO and NLO EW accuracy.
Afterwards, the following effects are discussed in detail: spin correlations between the $\tau$-leptons, finite-mass effects of the final-state leptons, NLO EW corrections, and the dependence on the parametrization of the photon flux.

\subsection{State-of-the-art Standard Model prediction}
\label{sec:Best_SM}

At LO and NLO EW, the fiducial cross sections read
\begin{align} \label{eq:best_SM}
    \mwm \mwm 
    \sigma^\LO  =  45.872(4) \, \mathrm{nb}, 
    \mwm \mwm \mwm \mwm \mwm \mwm \mwm \mwm \mwm 
    \sigma^\NLO =  45.331(4) \, \mathrm{nb}, 
    \mwm
\end{align}
with
\begin{align} 
    \Delta \sigma^\NLO  = \sigma^\NLO - \sigma^\LO = -0.542(1) \, \mathrm{nb}, 
    \mwm \mwm \mwm \,
    \delta = \frac{\Delta\sigma^\NLO}{\sigma^\LO} = -1.181(1) \, \% . 
    \,\,
\end{align}
These constitute our best predictions in the sense that they include spin-correlation effects and keep the full dependence on the masses of the final-state leptons. 
The photon flux is parameterized using the ChFF~\cite{Vidovic:1992ik} of the heavy ions,
combined with a non-trivial survival probability of the heavy ions 
during the collisions~\cite{Shao:2022cly}.

The relative size of the NLO EW correction to the fiducial cross section is $-1.2\,\%$ with respect to the LO contribution, consistent with the typical size of \order{\alpha} corrections.

\begin{figure}
    \centering{
    \raisebox{0pt}{\includegraphics[width=0.5\columnwidth]{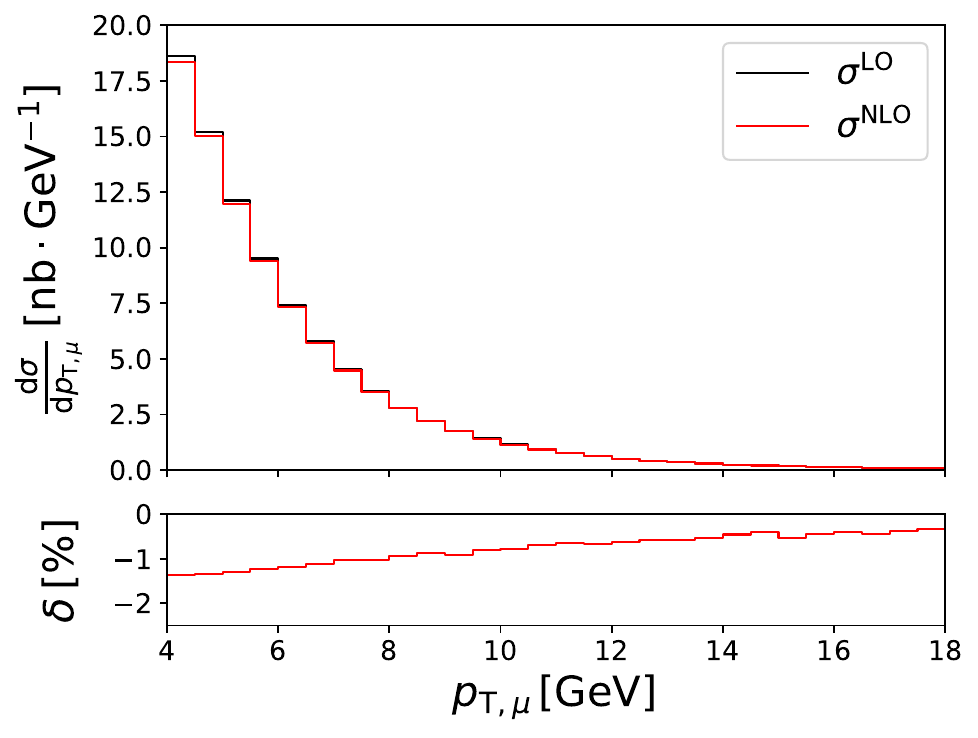}}
    \hspace{-0.3cm}
    \raisebox{0pt}{\includegraphics[width=0.5\columnwidth]{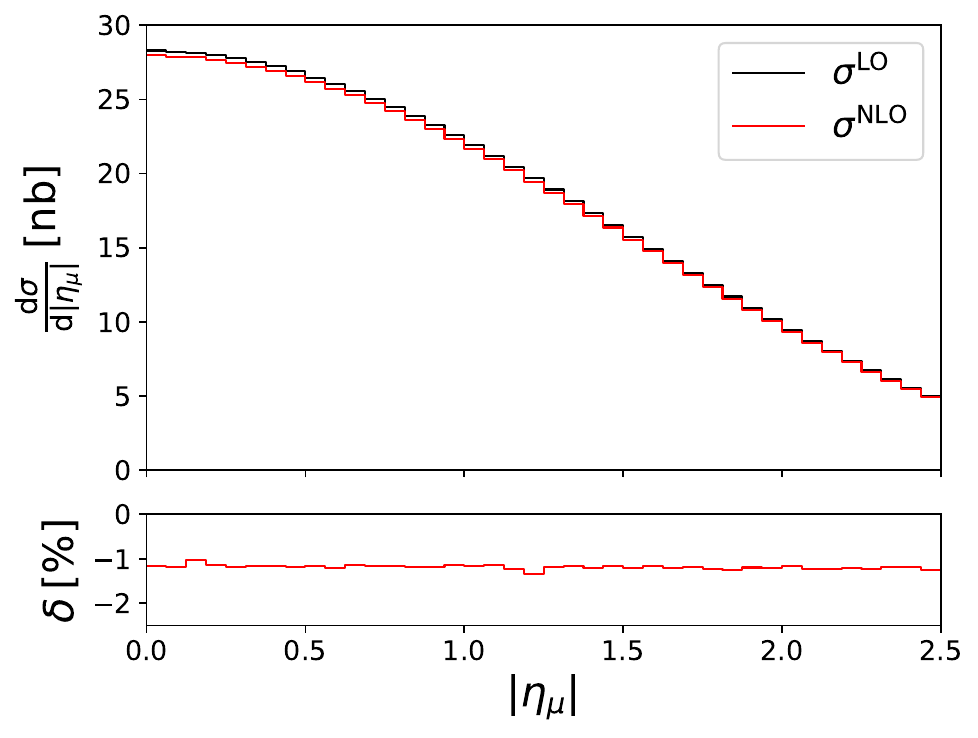}}
    \newline
    (a) \hspace{7.3cm} (b) \hspace{-5cm}
    \newline   
    \raisebox{0pt}{\includegraphics[width=0.5\columnwidth]{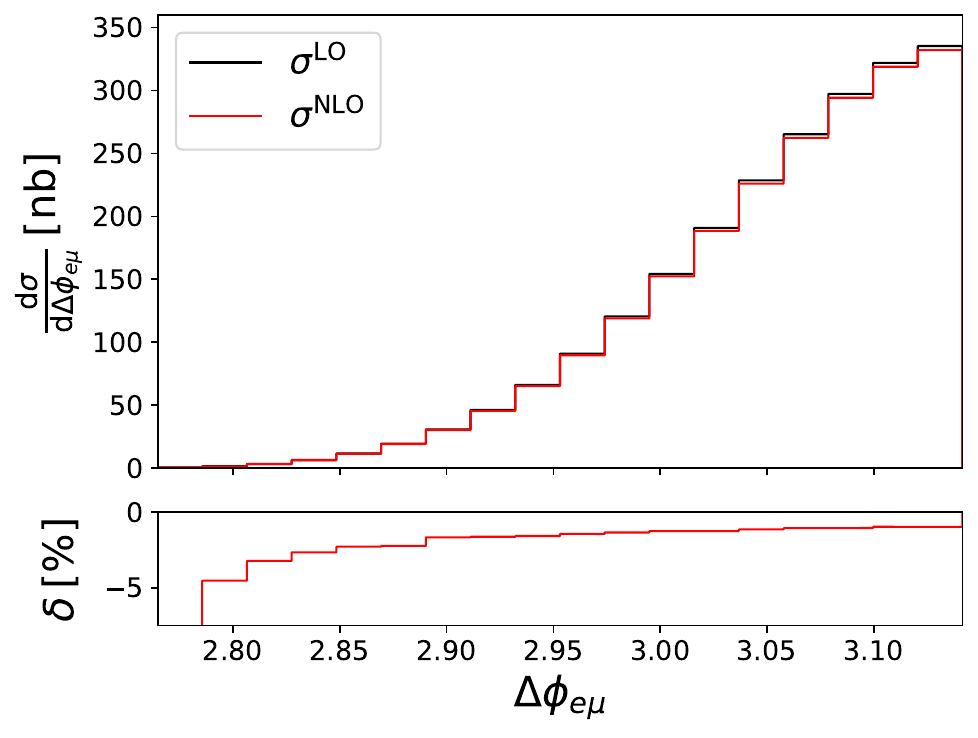}}
    \hspace{-0.3cm}
    \raisebox{0pt}{\includegraphics[width=0.5\columnwidth]{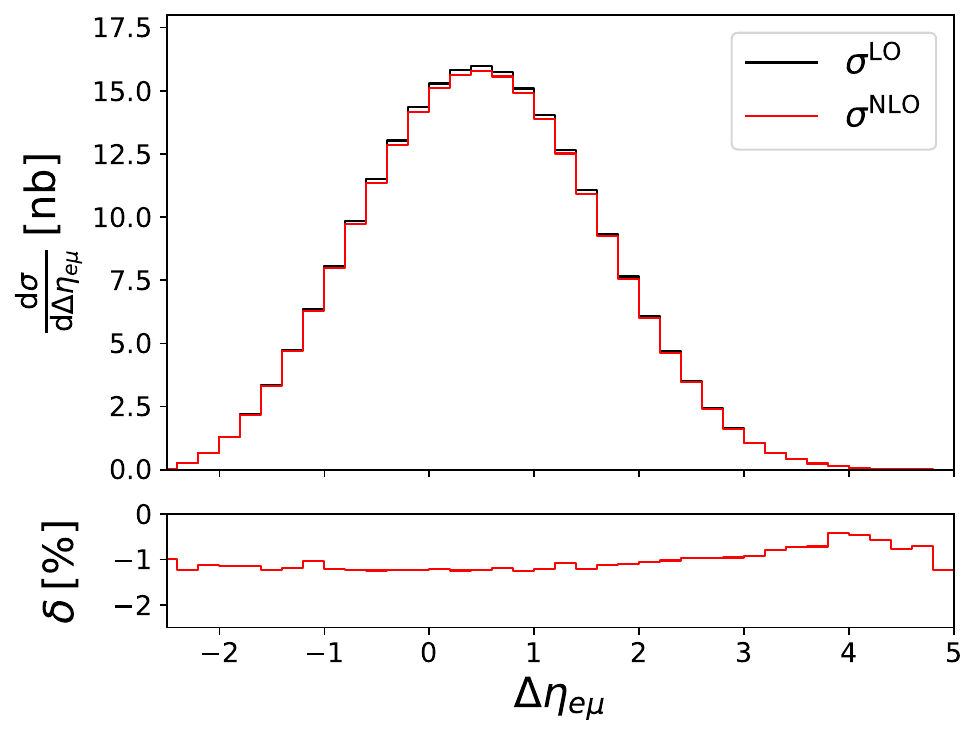}}
    \newline
    (c) \hspace{7.15cm} (d) \hspace{-1.4cm}
    }
    \caption{State-of-the-art SM prediction for $\processmue$ induced by UPCs of two lead ions at LO (black) and including NLO EW corrections (red) for:
    (a) the transverse momentum of the muon, 
    (b) the pseudorapidity of the muon,
    (c) the azimuthal angle distance between the muon and the positron, 
    and (d) the pseudorapidity difference between the muon and the positron.
    The lower panels show the relative NLO EW correction.
    }
    \label{fig:nlo_plots}
\end{figure}

At LO, the typical kinematical event configuration consists of low-energy muons concentrated in the central region of the detector, see Figs.~\ref{fig:nlo_plots}~(a),~(b). 
The inclusion of NLO EW corrections results in a uniform downshift of $-1.2\,\%$ in the pseudorapidity of the muon without any shape distortion.
On the other hand, the corrections slightly modify the shape of the $p_{\mathrm{T},\mu}$ distribution, with largest effect at low transverse momentum. 
The corrections are most negative for the smallest $p_{\mathrm{T},\mu}$ values allowed by the cut~(\ref{eq:lepton}), because hard photon radiation has the highest probability to push events below the $p_{\mathrm{T},\mu}$ cut in this region.
Note that there is no imprint of collinear enhancement due to final-state radiation, which would lead to larger negative corrections for large $p_{\mathrm{T},\mu}$, even after applying the photon-recombination procedure.
Since photon recombination is applied to both muons and positrons, there is no big difference between them in the final state. 
Therefore, only the muonic distributions are presented here.

Since the $\tau$-leptons are produced via a $2\to2$ kinematical configuration at LO, they are produced back-to-back.
This is reflected in the azimuthal angle distance between the muon and the positron, see Fig.~\ref{fig:nlo_plots}~(c). 
The particles originating from $\tau$-lepton decays receive a significant boost along the direction of the parent $\tau$-lepton, leading to a peak in the azimuthal angle distance between the muon and the positron at $\Delta \phi_{e \mu} = \pi$. 
This peak is smeared 
due to the presence of neutrinos in the decays of the $\tau$-leptons, but nearly all events concentrate in the back-to-back region of muon and positron where $\Delta \phi_{e \mu} > 2.8$.
The distribution for $\Delta \phi_{e \mu} < 2.75$ is not shown due to the smallness of the cross section there, despite the angular difference actually covering $\Delta \phi_{e \mu} \in [0,\pi]$.

Although the muon and positron preferentially exhibit a back-to-back configuration, the pseudorapidity difference peaks at $\Delta \eta_{e \mu} = 0.25$ rather than at $\Delta \eta_{e \mu} = 0$, see Fig.~\ref{fig:nlo_plots}~(d).  
This originates from our definition of the direction of the positive $z$-axis.
The inclusion of NLO EW corrections results in a $-1.2\,\%$ downshift across the whole phase space expect at high pseudorapidity difference, where the presence of real radiation becomes more pronounced.

\subsection{Spin-correlation effects}
\label{sec:spin_effects}


The spin-correlation effects of the produced $\tau$-leptons have an impact on the kinematics of the final-state particles and are therefore crucial for precise predictions. 
In general, spin-correlation effects can influence any observable where the kinematics of the decay products is not fully integrated over (see, {\it e.g.,} \cite{Dittmaier:2009un}).

\begin{figure}
    \centering{
    \raisebox{0pt}{\includegraphics[width=0.5\columnwidth]{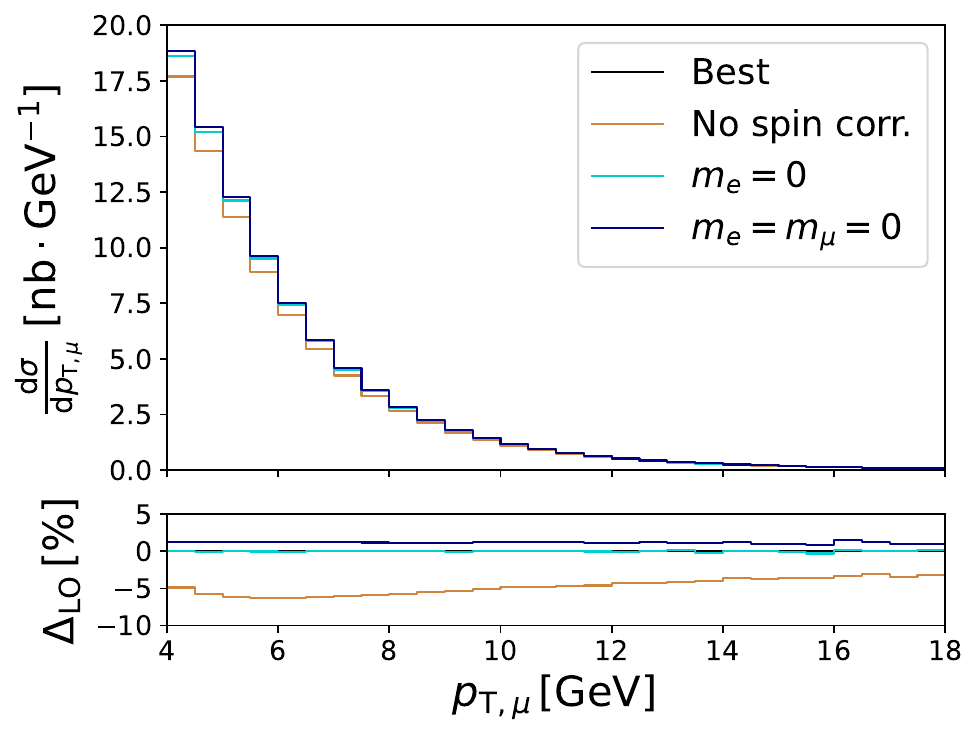}}
    \hspace{-0.3cm}
    \raisebox{0pt}{\includegraphics[width=0.5\columnwidth]{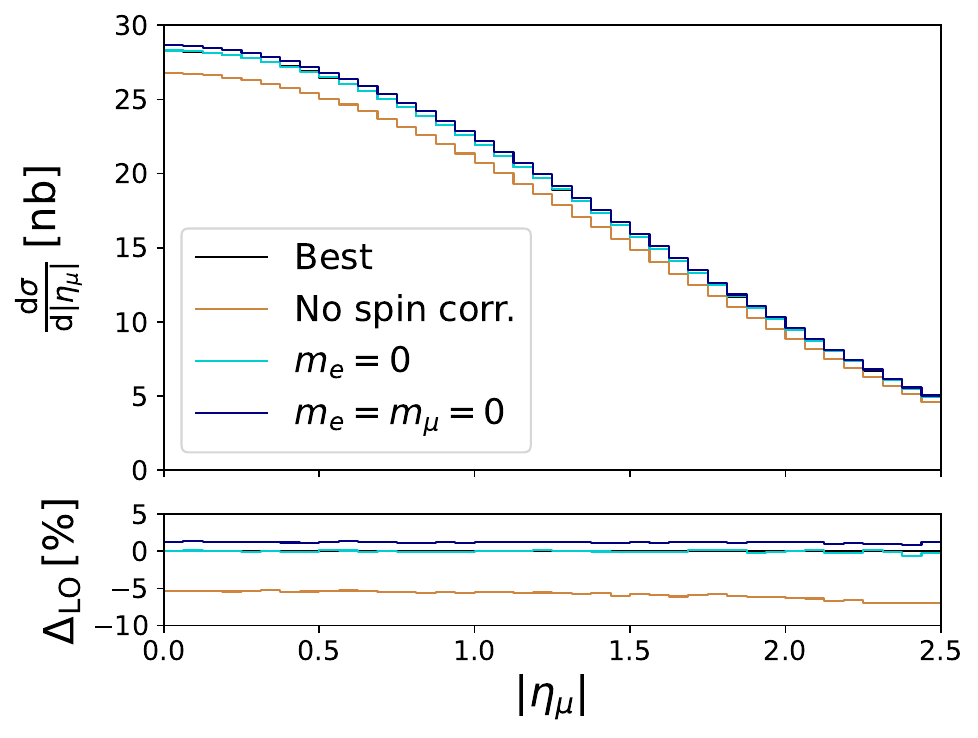}}
    \newline
    (a) \hspace{7.3cm} (b) \hspace{-5cm}
    \newline   
    \raisebox{0pt}{\includegraphics[width=0.5\columnwidth]{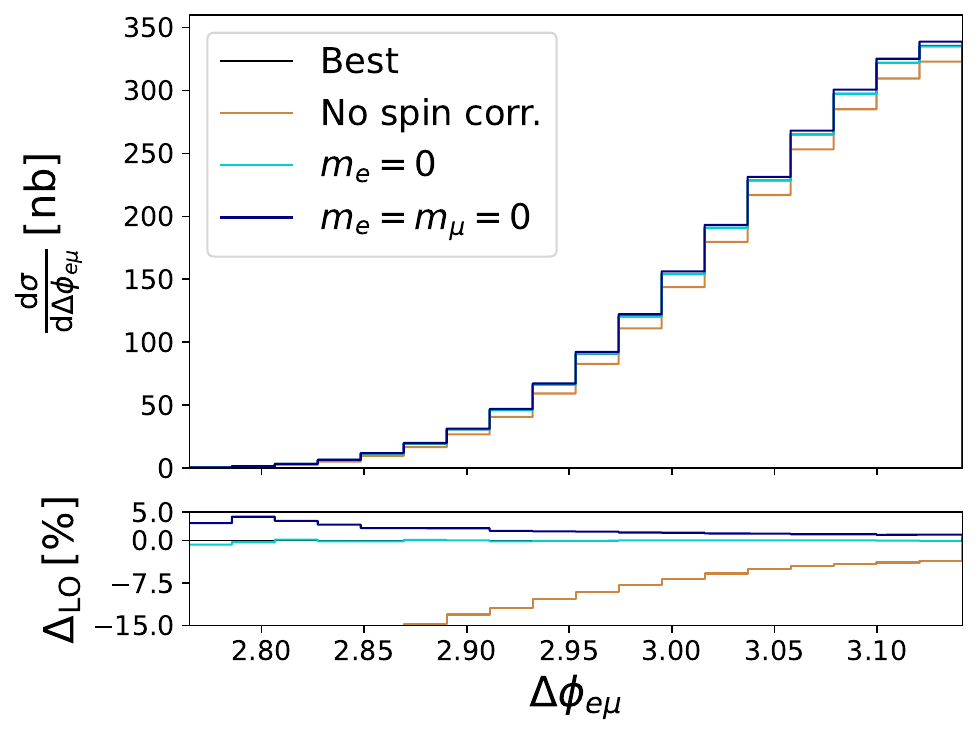}}
    \hspace{-0.3cm}
    \raisebox{0pt}{\includegraphics[width=0.5\columnwidth]{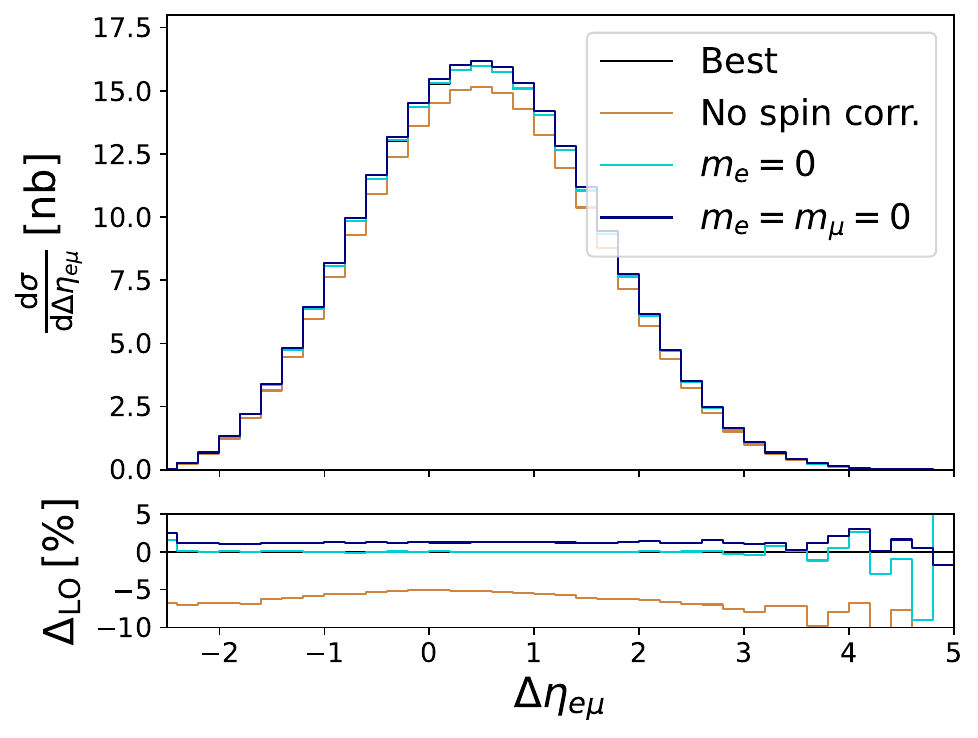}}
    \newline
    (c) \hspace{7.15cm} (d) \hspace{-1.4cm}
    }
    \caption{
    LO SM predictions for $\processmue$ induced by UPCs of two lead ions for:
    (a) the transverse momentum of the muon, 
    (b) the pseudorapidity of the muon,
    (c) the azimuthal angle distance between the muon and the positron, 
    and (d) the pseudorapidity difference between the muon and the positron.
    The predictions are calculated: 
    (black) using our best configuration, 
    (brown) neglecting spin-correlation effects, 
    (light-blue) considering the positron massless, 
    and (dark-blue) taking the positron and the muon massless.
    The lower panels show the relative deviation
  $\Delta_{\LO} = \sigma^\LO/\sigma^\LO_\mathrm{Best} - 1$. 
    }
    \label{fig:effects}
\end{figure}

In this section, we study the impact of spin correlations by comparing our best LO predictions with LO predictions that neglect the spin correlations between the $\tau$-leptons, \ie we compare the improved NWA with the naive NWA. 
The fiducial cross section predicted by the naive NWA is given in the third row of Table~\ref{tbl:effects}, while the corresponding differential distributions are shown in Fig.~\ref{fig:effects} (brown curves).
\begin{table}
  \centering{
    \begin{tabular}{c|c|c}
                        \hl & $\sigma^{\LO} [\text{nb}]$ & $\Delta_\LO \, [\%]$  \\
    \hline \hline
    Best                \hl & $45.872(4)$                & -                     \\
    \hline
    No spin corr.       \hl & $43.287(4)$                & $-5.64(1)$            \\
    \hline
    $m_e = 0$           \hl & $45.870(4)$                & $-0.01(1)$            \\
    \hline 
    $m_{\mu} = m_e = 0$ \hl & $46.444(4)$                & $\mwm1.23(1)$         \\
    \end{tabular}
  }
  \caption{
  LO fiducial cross section for $\processmue$ induced by UPCs of two lead ions. 
  The second row shows our best LO prediction. 
  All numbers but the ones in the third row include spin-correlation effects of the $\tau$-leptons.
  The fourth row gives the result assuming a massive muon and a massless electron, 
  while the fifth row considers both muon and positron as massless.
  The cross section is given in the second column and the relative deviation
  $\Delta_{\LO} = \sigma^\LO/\sigma^\LO_\mathrm{Best} - 1$  
  is provided in the third column.
  }
  \label{tbl:effects}
\end{table}

For the fiducial cross section, the spin-correlations effects are found to be
more than $5\,\%$, proving the necessity of their inclusion in phenomenological studies.
The effect is rather uniform in distributions and largest at low transverse momentum and/or large pseudorapidity of the muon or positron.
For the azimuthal-angle difference between the two leptons, the spin correlations are minimal in the back-to-back region and increase further away from it.
The effect is rather uniform in $\Delta\eta_{e\mu}$ and somewhat more pronounced for the minimal and the maximal values of the difference in the pseudorapidity of the two leptons.

Finally, we would like to point out that the effect of spin correlations is of the same order as (and even larger than) the NLO EW corrections. 
Their inclusion is thus crucial for the precise study of $\tau$-pair production in UPCs.
Moreover, this effect is significantly larger (two orders of magnitude) than the $\sim0.05\,\%$ estimate given in App.~A of Ref.~\cite{Dyndal:2020yen}, where the process $\gamma\gamma\to\tau^+\tau^-\to\mu^+\mu^-\bar{\nu}_{\tau}\nu_{\tau}\nu_{\mu}\bar{\nu}_{\mu}$ is considered for a fixed photon--photon centre-of-mass energy.

\subsection{Effects of non-zero lepton masses}
\label{sec:mass_effects}



The effect of neglecting lepton masses can be estimated by the ratio between the mass $m_{\ell}$ and the hard energy scale $\sqrt{s_{\gamma\gamma}}$ of the process, \emph{i.e.}\ by $m_{\ell}/\sqrt{s_{\gamma\gamma}}$.
Neglecting lepton masses in high-energy proton--proton collisions at the LHC, where $\sqrt{s_{\gamma\gamma}} \gsim 100 \, \GeV$, is hence a very good approximation.
For the investigated process, however, the energy scale is of order the mass $m_{\tau}$ of the $\tau$-lepton. 
Thus, if the final-state charged lepton is considered massless, the neglected effects are expected to be of order $m_{\ell}/m_{\tau}$. 
For this reason, the massless approximation is presumably good for final-state positrons but not necessarily for final-state muons.


In this section, we study the impact of non-zero lepton masses by comparing our best LO predictions with a LO prediction that neglects the electron mass and with LO predictions 
based on the approximations $m_{\mathrm{e}}=0$ or $m_{\mathrm{e}}=m_\mu=0$.
The fiducial cross section predicted for massless electrons is given in the fourth row of Table~\ref{tbl:effects}, while the corresponding differential distributions are shown in Fig.~\ref{fig:effects} (light-blue curves). 
Moreover, the fiducial cross section predicted for massless positrons and muons is provided in the fifth row of Table~\ref{tbl:effects}, while the corresponding differential distributions are illustrated in Fig.~\ref{fig:effects} (dark-blue curves).

As anticipated, considering massless positrons is a valid approximation since the error made at the level of the fiducial cross section and for the differential distributions is 
even smaller than the Monte Carlo integration error. 
On the other hand, the error made in the massless-muon approximation is at the level of 
$1.2\,\%$ with respect to our best LO prediction.
Therefore, retaining the dependence on the muon mass is important for obtaining precise predictions for the considered process.
At the level of the differential distributions, 
the effect essentially results in an upwards shift, and no shape distortion is observed in the relevant part of phase space.

\subsection{Next-to-leading-order electroweak effects}
\label{sec:NLO_effects}

In the results presented in Sec.~\ref{sec:Best_SM}, the full NLO EW correction to the considered process was included.
In order to study the EW corrections in more detail, we divide the NLO EW correction into several gauge-invariant subsets, as discussed in Sec.~\ref{sec:NLO}.
We also study the effects of a non-inclusive treatment of collinear radiation.

\subsubsection*{Various types of EW corrections}

In Table~\ref{tbl:NLO}, we present the absolute and relative values of the subcontributions to the NLO EW corrections at the level of the fiducial cross section.
\begin{table}
  \centering{
    \begin{tabular}{c|c|c|c}
    subprocess                                               & correction \hl & $\Delta \sigma^{\NLO} \, [\text{nb}]$    & $\delta \, [\%]$ \\
    \hline \hline
    \multirow{4}{*}{$\gamma\gamma\to\tau^+\tau^-$}           & QED        \hl & \wm$0.1737(3)$\wn                        & \wm$0.3784(6)$   \\
    \cline{2-4}
                                                             & weak       \hl & \wm$0.00082(1)$                          & \wm$0.0018(1)$   \\
    \cline{2-4}
                                                             & fermionic  \hl & \wm$0.00005(1)$                          & \wm$0.0001(1)$   \\
    \hhline{~===} 
                                                             & sum        \hl & \wm$0.1745(3)$\wn                        & \wm$0.3803(6)$   \\
    \hline \hline
    \multirow{3}{*}{$\tau^-\to\mu^-\nu_{\tau}\bar{\nu}_\mu$} & bosonic    \hl & $-0.3342(2)$\wn                          & $-0.7287(5)$     \\
    \cline{2-4}
                                                             & fermionic  \hl & \wm$0.0010(1)$\wn                        & \wm$0.0023(1)$   \\
    \hhline{~===} 
                                                             & sum        \hl & $-0.3332(2)$\wn                          & $-0.7264(4)$     \\
    \hline \hline                                                     
    \multirow{3}{*}{$\tau^+\to e^+\bar{\nu}_{\tau}\nu_e$}    & bosonic    \hl & $-0.3838(3)$\wn                          & $-0.8368(7)$     \\
    \cline{2-4}                                                     
                                                             & fermionic  \hl & \wm$0.0010(1)$\wn                        & \wm$0.0023(1)$   \\
    \hhline{~===} 
                                                             & sum        \hl & $-0.3828(4)$\wn                          & $-0.8345(7)$     \\
    \hline \hline
    \multicolumn{2}{c|}{sum \hl}                                              & $-0.5416(5)$\wn                          & $-1.181(1)$\wn 
    \end{tabular}
  }
  \caption{
  Subcontributions of NLO EW corrections to $\processmue$ induced by UPCs of two lead ions. 
  The three blocks of results give the corrections to $\tau$-pair production, $\tau^-$-lepton decay, and $\tau^+$ decay, respectively.
  The last row gives the total NLO EW correction as the sum of these three contributions.
  The corrections to $\tau$-pair production are split into the QED, weak, and fermionic corrections, 
  and the corrections to $\tau$-leptons decays are divided into bosonic and fermionic corrections.
  The absolute corrections are shown in the third column and the relative corrections are contained in the last column.
  }
  \label{tbl:NLO}
\end{table}
For the three subprocesses the relative corrections to the fiducial cross sections are generically at the percent level or somewhat below, which is typical for \order{\alpha} corrections without particular enhancement. 
However, while the correction to $\tau$-pair production is positive, the corrections to $\tau$-lepton decays are negative and two times larger than the corrections to $\tau$-pair production. 
The corrections due to closed fermion loops to each subprocesses are found to be the smallest and phenomenologically insignificant.
This is not surprising because the input-parameter scheme for the electromagnetic coupling has been chosen such that the dominant part of the closed fermion-loop contributions is absorbed into the definition of the electromagnetic coupling.
Note that the correction to the leptonic $\tau$-decays from closed fermion loops does not depend on the type of external charged lepton and, thus, it is found that the fermionic contributions are of the same size for both $\tau$-lepton decays.
However,  
the bosonic corrections, which include QED final-state radiation as well as 
loops involving W, Z, and/or Higgs bosons (see Sec.~\ref{sec:NLO}),
mildly depend on the mass of the final-state charged lepton.
This is reflected by the larger negative correction for the $\tau^+$-lepton decay into a positron. 
Finally, we find that in the bosonic corrections to $\tau$-pair production, \ie weak and QED corrections, the weak contribution turns out to be suppressed with respect to the QED contribution.
The suppression factor roughly resembles $\bar{s}_{\gamma\gamma}/M^2_\mathrm{W}$ with $\sqrt{ \vphantom{|}\bar{s}_{\gamma\gamma}}$ representing a typical hard scattering energy of a few times $2m_\tau$.

\begin{figure}
    \centering{
    \raisebox{0pt}{\includegraphics[width=0.5\columnwidth]{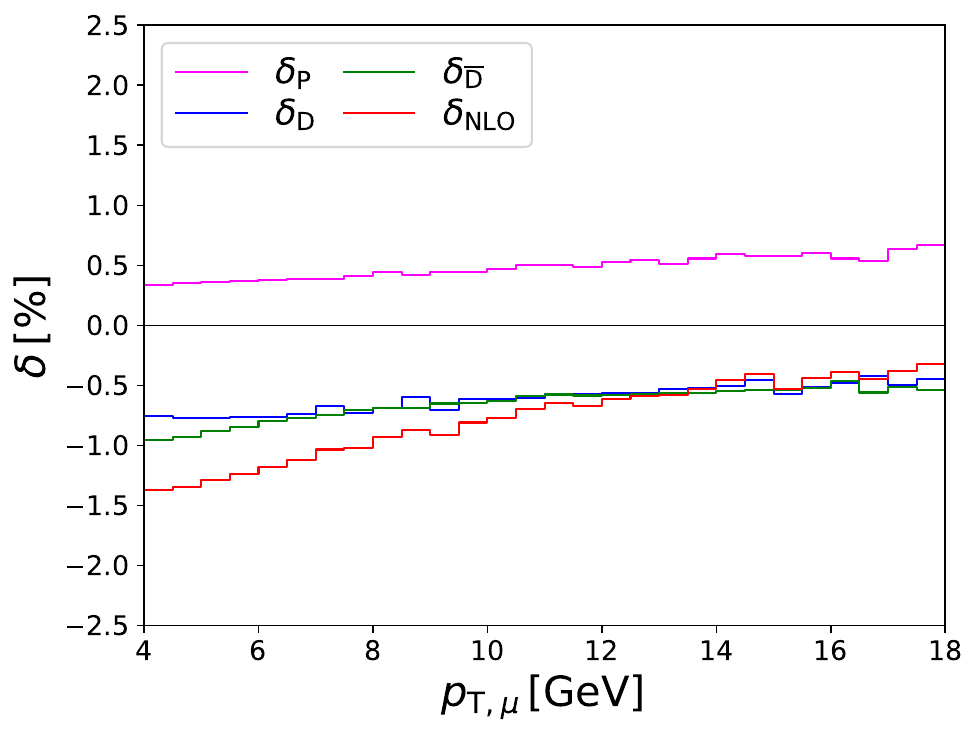}}
    \hspace{-0.3cm}
    \raisebox{0pt}{\includegraphics[width=0.5\columnwidth]{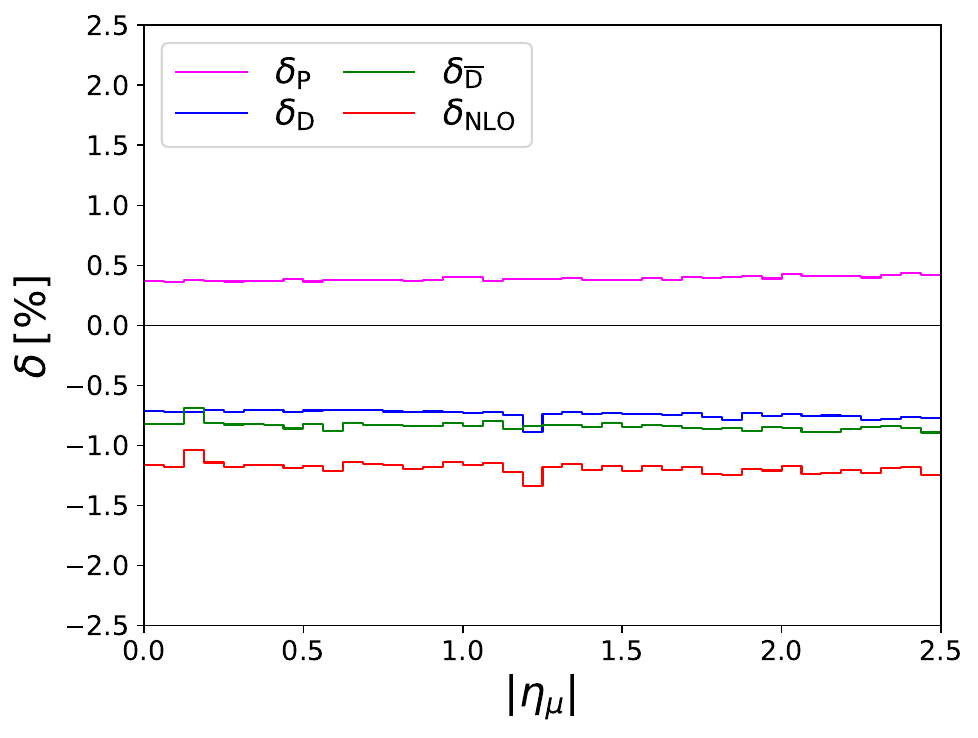}}
    \newline
    (a) \hspace{7.3cm} (b) \hspace{-5cm}
    \newline
    \raisebox{0pt}{\includegraphics[width=0.5\columnwidth]{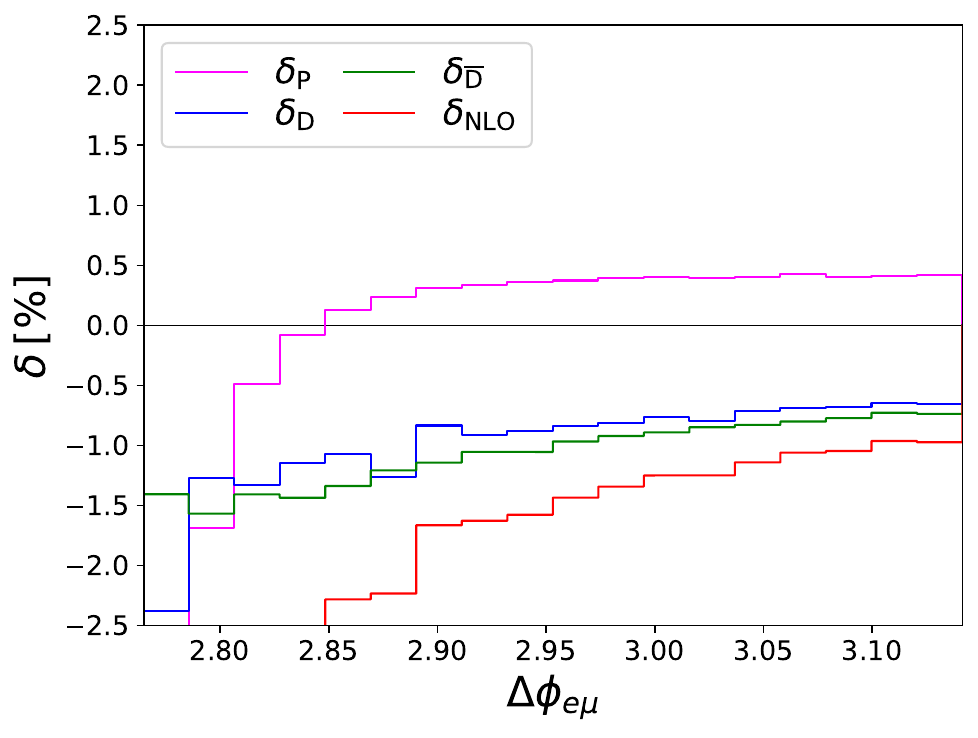}}
    \hspace{-0.3cm}
    \raisebox{0pt}{\includegraphics[width=0.5\columnwidth]{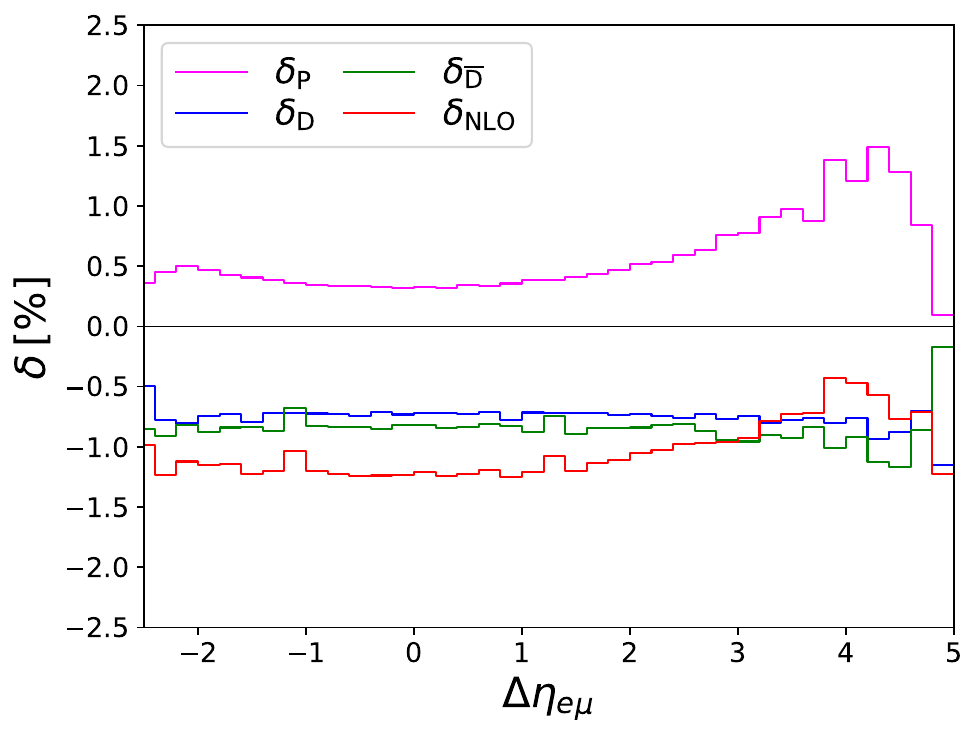}}
    \newline
    (c) \hspace{7.2cm} (d) \hspace{-1.3cm}
    }
    \caption{
    Relative NLO EW corrections to $\processmue$ induced by UPCs of two lead ions for:
    (a) the transverse momentum of the muon, 
    (b) the pseudorapidity of the muon,
    (c) the azimuthal angle distance between the muon and the positron, 
    and (d) the pseudorapidity difference between the muon and the positron.
    The predictions are split into corrections to: 
    (pink) $\tau$-pair production, 
    (blue) $\tau^-$-lepton decay, 
    and (green) $\tau^+$-lepton decay. 
    The total relative EW correction is provided in red.
    }
    \label{fig:NLO_split}
\end{figure}

In Fig.~\ref{fig:NLO_split}, we present the relative NLO EW correction to differential distributions split according to subprocesses. 
The dominant contributions arise from QED corrections to $\tau$-pair production and from bosonic corrections to the $\tau$-lepton decays.  
The corrections to each subprocess do not exhibit significant shape distortions in the relevant phase-space regions.
However, for the corrections to the transverse momentum of the muon, a slight slope is observed: The corrections to $\tau$-lepton decays show a steeper, negative trend, whereas the corrections to $\tau$-pair production are positive.
For the azimuthal angle difference, the largest correction in the tail of the distribution arises from the corrections to $\tau$-pair production.
The corrections to the $\tau^+$ and $\tau^-$ decays are almost identical and only differ due to the difference in the muon and electron masses.

\subsubsection*{Effects of a non-inclusive treatment of collinear radiation}

In precision calculations for processes with light charged particles, the treatment of final-state radiation requires particular care.
The Kinoshita--Lee--Nauenberg (KLN) theorem~\cite{Kinoshita:1962ur, Lee:1964is} states that, as a consequence of unitarity, the IR singularities originating from 
virtual corrections completely cancel against the IR singularities arising from the phase-space integral in the evaluation of real-emission corrections, provided that the observable is inclusive with respect to soft or collinear photon radiation. 
However, if real radiation is not treated inclusively in the collinear limit, the terms proportional to $\alpha \ln{m_f}$ arising from the collinear emission of a photon do not fully cancel the $\alpha \ln{m_f}$ terms from virtual corrections in observables that depend on the energy splitting between the emitter and the radiated photon. 
Such observables are known as non-collinear-safe observables.\footnotemark
\footnotetext{
Note that while the remaining terms proportional to $\alpha \ln{m_\mu}$ are formally finite, they become potentially large when typical energy scales $Q$ significantly exceed the muon mass, even diverging in the limit $m_\mu/Q\to0$.
For this reason, such terms are commonly classified as singular terms in higher-order calculations for high-energy collisions.
}
In this section, we examine the effect of a non-inclusive treatment of collinear radiation off the muon, \ie when no muon--photon recombination procedure is applied.
Note that a positron--photon recombination procedure is still performed, as collinear radiation off $e^\pm$ cannot be individually resolved by the electromagnetic calorimeters of the detectors at the LHC.

We present the NLO EW corrections for $\processmue$ using either the inclusive, $\Delta\sigma_\drs^\NLO$, or the non-inclusive, $\Delta\sigma_\bare^\NLO$, treatment of collinear radiation off muons. 
The NLO EW effects are provided as absolute corrections and as relative corrections with respect to the LO prediction, 
\begin{align}
  \delta_i = \frac{\Delta\sigma_i^\NLO}{\sigma^\LO},
\end{align}
where the subscript $i = \drs, \bare$, indicates whether muons are dressed or not, respectively. 
Additionally, we show the difference between the relative corrections to the two predictions,
\begin{align}
  \Delta_{\drs/\bare} = \frac{\sigma_\drs^\NLO - \sigma_\bare^\NLO}{\sigma^\LO}
 = \delta_\drs - \delta_\bare.
\label{eq:delta_drsbare}
\end{align}

In Table~\ref{tbl:NLO_non_safe_massive}, we present the NLO EW corrections for dressed and bare muons retaining the full mass dependence of all leptons.
\begin{table}
  \centering{
    \begin{tabular}{c||c|c||c|c||c}
    subprocess \hl & $\Delta \sigma^{\NLO}_\bare \,[\text{nb}]$ & $\delta_\bare \, [\%]$ & $\Delta \sigma^{\NLO}_\drs\,[\text{nb}]$ & $\delta_\drs\, [\%]$ & $\Delta_{\drs/\bare} \, [\%]$ \\
    \hline \hline
     $\gamma\gamma\to\tau^+\tau^-$           \hl & \wm$0.1666(3)$ & \wm$0.363(1)$ & \wm$0.1745(3)$ & \wm$0.380(1)$ & \wm$0.017(1)$ \\
    \hline
    $\tau^-\to\mu^-\nu_{\tau}\bar{\nu}_\mu$  \hl & $-0.4799(2)$   & $-1.046(1)$   & $-0.3332(2)$   & $-0.726(1)$   & \wm$0.320(1)$ \\
    \hline
    $\tau^+\to e^+\bar{\nu}_{\tau}\nu_e$     \hl & $-0.3821(3)$   & $-0.833(1)$   & $-0.3828(3)$   & $-0.835(1)$   & $-0.002(1)$   \\
    \hline\hline
    sum                                      \hl & $-0.6954(5)$   & $-1.516(1)$   & $-0.5417(5)$   & $-1.181(1)$   & \wm$0.335(2)$    
    \end{tabular}
  }
  \caption{
  NLO EW corrections for $\processmue$ induced by UPCs of two lead ions for massive muons and positrons.
  The second, third, and fourth rows present the corrections to $\tau$-pair production, $\tau^-$-lepton decay, and $\tau^+$-lepton decay, respectively.
  The last row gives the total NLO EW correction as the sum of these three contributions.
  The absolute and relative corrections obtained for bare muons are shown in the second and third columns, respectively.
  The fourth and fifth columns provide the absolute and relative corrections for dressed muons.
  The final column presents the difference (\ref{eq:delta_drsbare})
between the relative EW corrections computed for dressed and bare muons, normalized by the LO prediction.
  }
  \label{tbl:NLO_non_safe_massive}
\end{table}
As discussed in Sec.~\ref{sec:mass_effects}, the muon mass is not negligible compared to the energy scale of the process under consideration. 
Thus, the uncancelled term proportional to $\alpha \ln{m_{\mu}}$ do not lead to particularly large corrections in the prediction for bare muons. 
Nevertheless, the muon--photon recombination has an impact on the prediction for the integrated NLO EW correction, mainly originating from the definition of the fiducial phase space, which involves a cut on the transverse momentum of the muon.
When photon recombination is employed for the muon, the absolute value of the total NLO EW correction decreases by approximately $0.3\,\%$ compared to the prediction for bare muons. 
This effect mainly arises from the NLO EW corrections to the $\tau^-$-lepton decay, where the photons can be emitted by the muon.
Note that the corrections to $\tau$-pair production and to $\tau^+$-lepton decay can still be slightly affected by the recombination, because the emitted photon can be collinear to the muon.

\begin{figure}
    \centering{
    \raisebox{0pt}{\includegraphics[width=0.5\columnwidth]{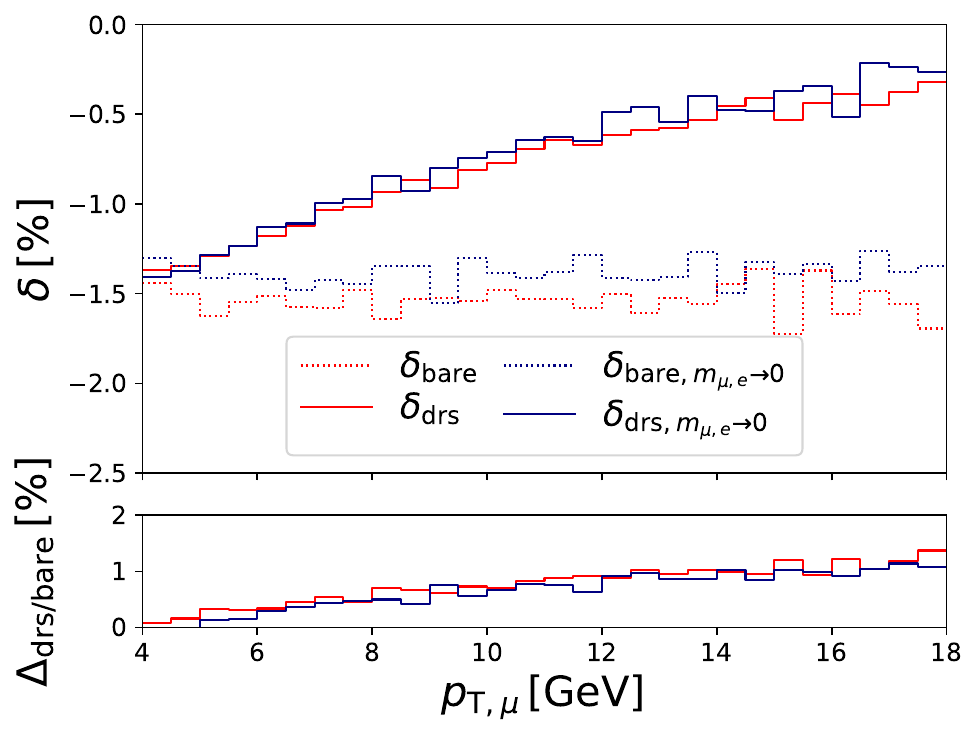}}
    \hspace{-0.3cm}
    \raisebox{0pt}{\includegraphics[width=0.5\columnwidth]{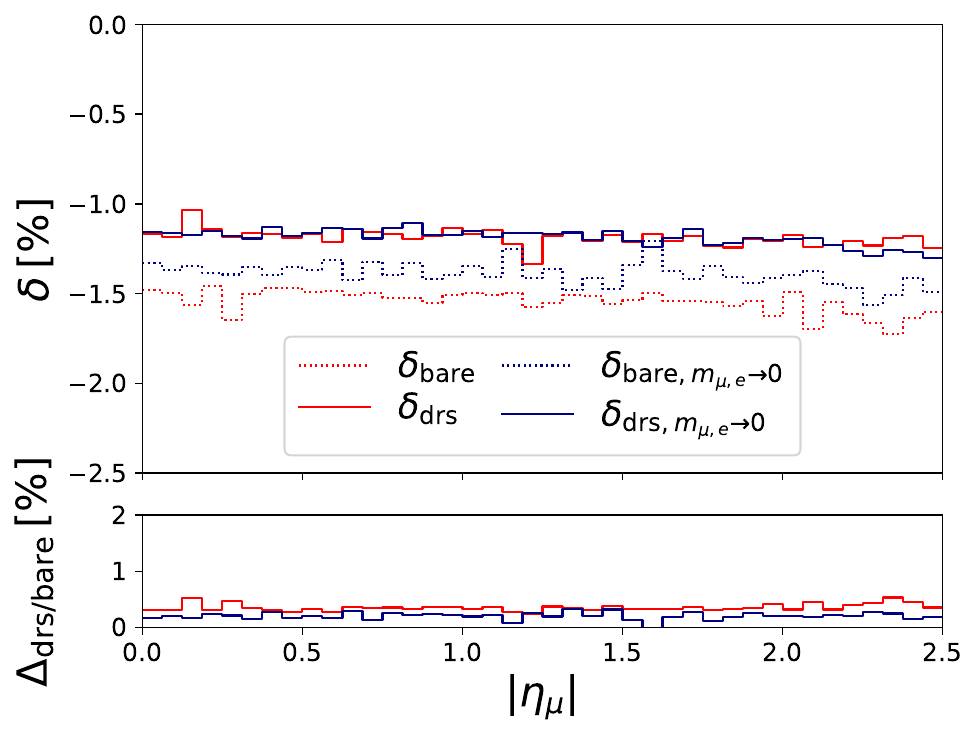}}
    \newline
    (a) \hspace{7.3cm} (b) \hspace{-5cm}
    \newline
    \raisebox{0pt}{\includegraphics[width=0.5\columnwidth]{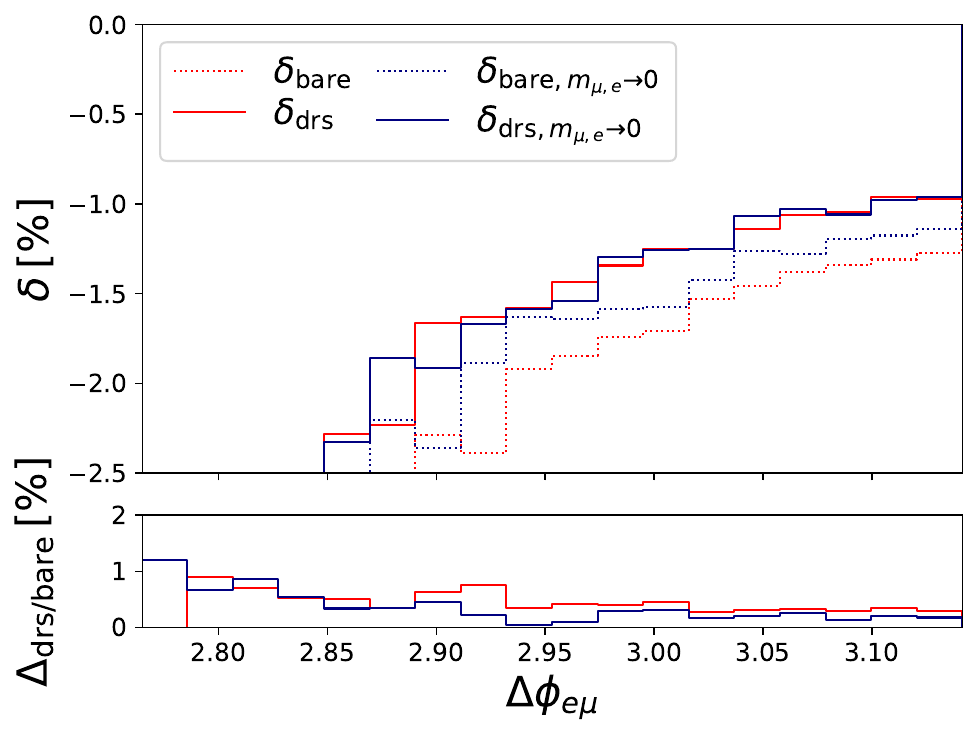}}
    \hspace{-0.3cm}
    \raisebox{0pt}{\includegraphics[width=0.5\columnwidth]{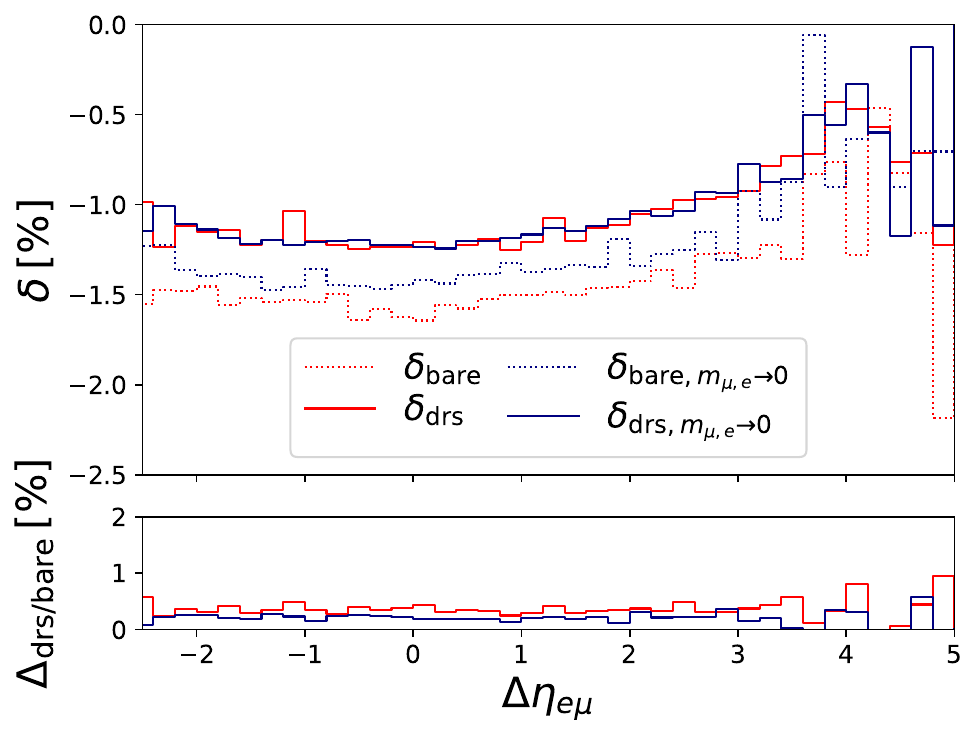}}
    \newline
    (c) \hspace{7.2cm} (d) \hspace{-1.3cm}
    }
    \caption{Relative NLO EW corrections for $\processmue$ induced by UPCs of two lead ions for:
    (a) the transverse momentum of the muon, 
    (b) the pseudorapidity of the muon,
    (c) the azimuthal angle distance between the muon and the positron, 
    and (d) the pseudorapidity difference between the muon and the positron.
    Predictions are computed using the physical (on-shell) masses of the final-state particles (red) and in the massless limit (blue).
    For each case, results are shown with (solid) and without (dotted) muon--photon recombination.
    The lower panels show the difference between relative corrections with and without muon--photon recombination, normalized by the corresponding LO prediction.
    }
    \label{fig:NLO_non_safe}
\end{figure}

In Fig.~\ref{fig:NLO_non_safe}, the relative NLO EW correction with respect to the LO prediction, for massive muons and positrons, is shown at the differential level for bare muons (solid red curves) and dressed muons (dashed red curves).
In the lower panels, the difference between the relative corrections for bare and dressed muons is provided (solid red line).
For observables that are not sensitive to the energy splitting between the muon and the photon in the collinear limit, such as $|\eta_{\mu}|$, $\Delta\phi_{e \mu}$, and $\Delta\eta_{e \mu}$, applying a muon–photon recombination procedure results in a uniform upward shift of the differential distributions of the same size as found in the corrections to the integrated cross section.
However, for $p_{\mathrm{T},\mu}$, which is a non-collinear-safe observable, the recombination procedure affects the shape of the distribution.
For bare muons, the relative corrections exhibit a nearly flat behaviour, indicating that, within the energy range of the considered process, collinear radiation off the muon is not particularly enhanced.

We extend the previous comparison to the case where the massless limit is taken for both the muon and the positron.
In this limit, uncancelled terms proportional to $\alpha \ln{m_{\mu}}$ ($m_\mu\to0$) remain in the correction, the treatment of which requires particular attention.
To consistently calculate the NLO EW correction for bare muons in the massless limit, we employ the extended version of the dipole subtraction formalism presented in Ref.~\cite{Dittmaier:2008md}, which allows for a non-inclusive treatment of collinear radiation off light fermions.
This formalism allows us to consistently retain the terms proportional to $\alpha \ln{m_\mu}$ and to evaluate them using the physical muon mass,
in spite of the use of massless muons in matrix elements.

The NLO EW corrections in the massless limit are provided in Table~\ref{tbl:NLO_non_safe_massless} at the level of the fiducial cross section and at differential level in Fig.~\ref{fig:NLO_non_safe} (blue curves).
\begin{table}
  \centering{
    \begin{tabular}{c||c|c||c|c||c}
    subprocess \hl  & $\Delta \sigma^{\NLO}_\bare \,[\text{nb}]$ & $\delta_\bare \, [\%]$ & $\Delta \sigma^{\NLO}_\drs \, [\text{nb}]$ & $\delta_\drs \, [\%]$ & $\Delta_{\drs/\bare}\, [\%]$   \\
    \hline \hline
    $\gamma\gamma\to\tau^+\tau^-$           \hl & \wm$0.1685(2)$ & \wm$0.363(1)$ & \wm$0.1762(2)$ & \wm$0.379(1)$ & $0.017(1)$ \\
    \hline
    $\tau^-\to\mu^-\nu_{\tau}\bar{\nu}_\mu$ \hl & $-0.4531(3)$   & $-0.976(1)$   & $-0.3595(3)$   & $-0.774(1)$   & $0.202(1)$ \\
    \hline
    $\tau^+\to e^+\bar{\nu}_{\tau}\nu_e$    \hl & $-0.3594(3)$   & $-0.774(1)$   & $-0.3597(3)$   & $-0.774(1)$   & $0.001(1)$ \\
    \hline \hline
    sum                                     \hl & $-0.6440(5)$   & $-1.387(1)$   & $-0.5429(5)$   & $-1.169(1)$   & $0.218(2)$      
    \end{tabular}
  }
  \caption{
  Same as Table~\ref{tbl:NLO_non_safe_massive}, but taking the massless limit for positrons and muons.
  }
  \label{tbl:NLO_non_safe_massless}
\end{table}
Note that the relative corrections are normalized to the LO prediction in the massless limit, which is given in the last row of Table~\ref{tbl:effects}.
As in the massive case, the total NLO EW correction for dressed muons is reduced with respect to the 
one for bare leptons. 
This reduction appears as a $0.2\,\%$ relative difference in the fiducial cross section and as a uniform downward shift of the same order in collinear-safe observables at the differential level.
In the transverse momentum distribution of the muon the same redistribution of events as the one described for the massive case is found.
As before, since the muon mass is not small relative to the considered energy range, no collinear enhancement is observed for bare muons from the evaluation of the terms proportional to $\alpha\ln{m_\mu}$; instead, the relative correction is uniformly distributed.

\subsection{Different models for the parametrization of the photon flux}
\label{sec:ph_flux_effects}


The parametrization of the photon flux is the largest source of uncertainty in predictions for UPCs.
State-of-the-art parametrizations use the charge form factor (ChFF)~\cite{Vidovic:1992ik} of the ions to characterize the photon density, \emph{i.e.}, the probability of radiating a photon with a given energy and impact parameter.
An alternative method employs the electric dipole form factor (EDFF)~\cite{Cahn:1990jk} instead. 
However, the ChFF-based approach is supposed to be more accurate, as the EDFF fails to describe the photon density at impact parameters smaller than the ion radius~\cite{Shao:2022cly}.

\begin{figure}
    \centering{
    \raisebox{0pt}{\includegraphics[width=0.5\columnwidth]{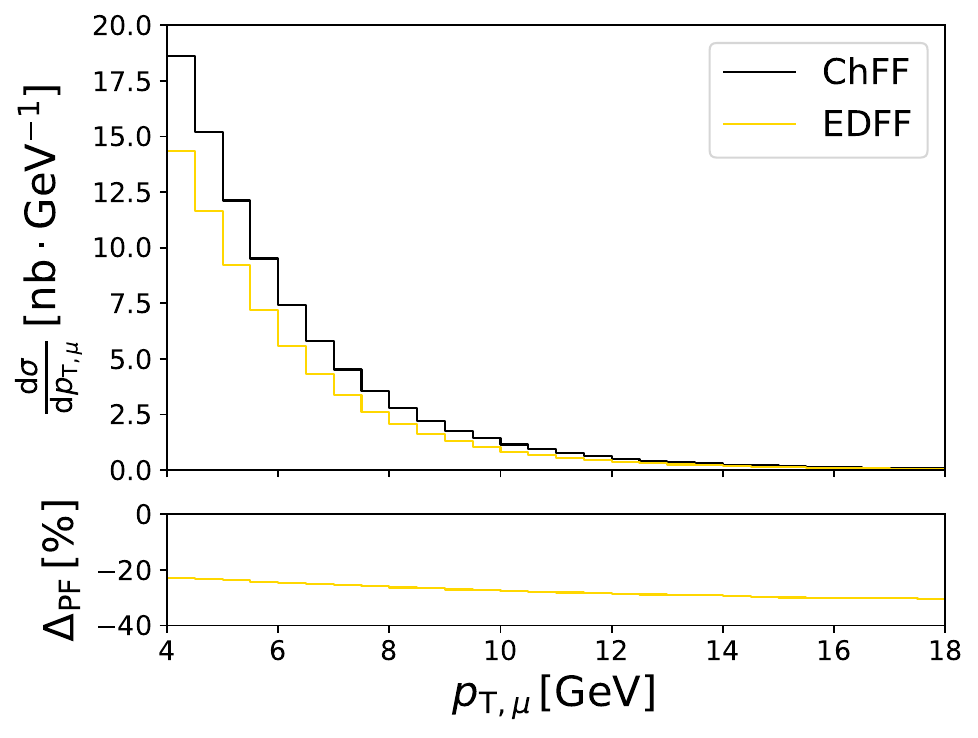}}
    \hspace{-0.3cm}
    \raisebox{0pt}{\includegraphics[width=0.5\columnwidth]{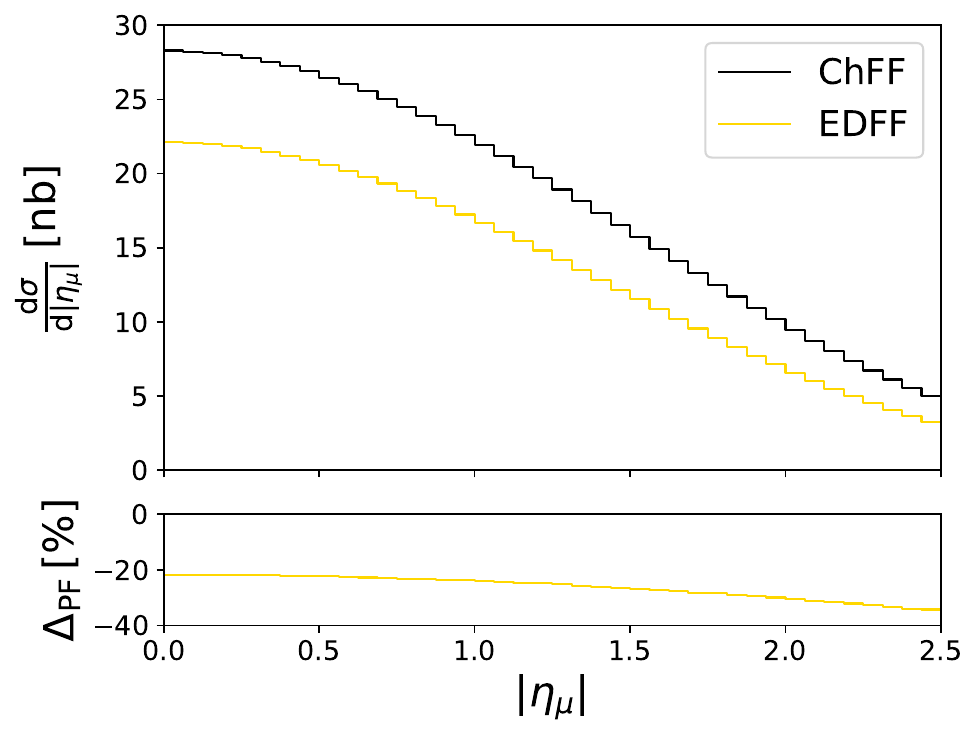}}
    \newline
    (a) \hspace{7.3cm} (b) \hspace{-5cm}
    \newline
    \raisebox{0pt}{\includegraphics[width=0.5\columnwidth]{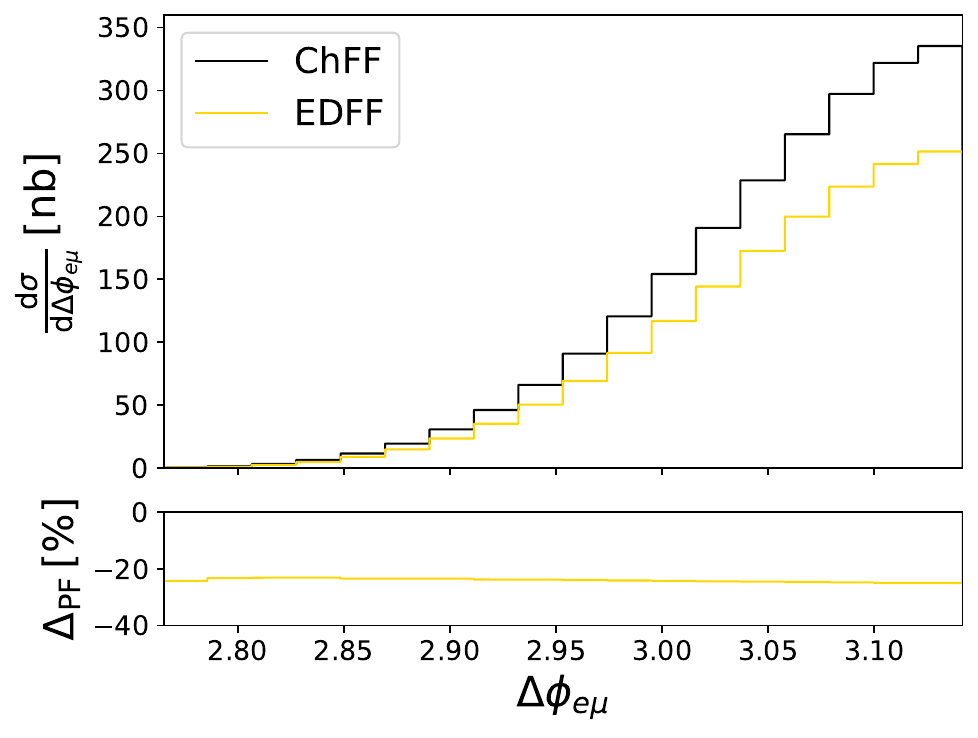}}
    \hspace{-0.3cm}
    \raisebox{0pt}{\includegraphics[width=0.5\columnwidth]{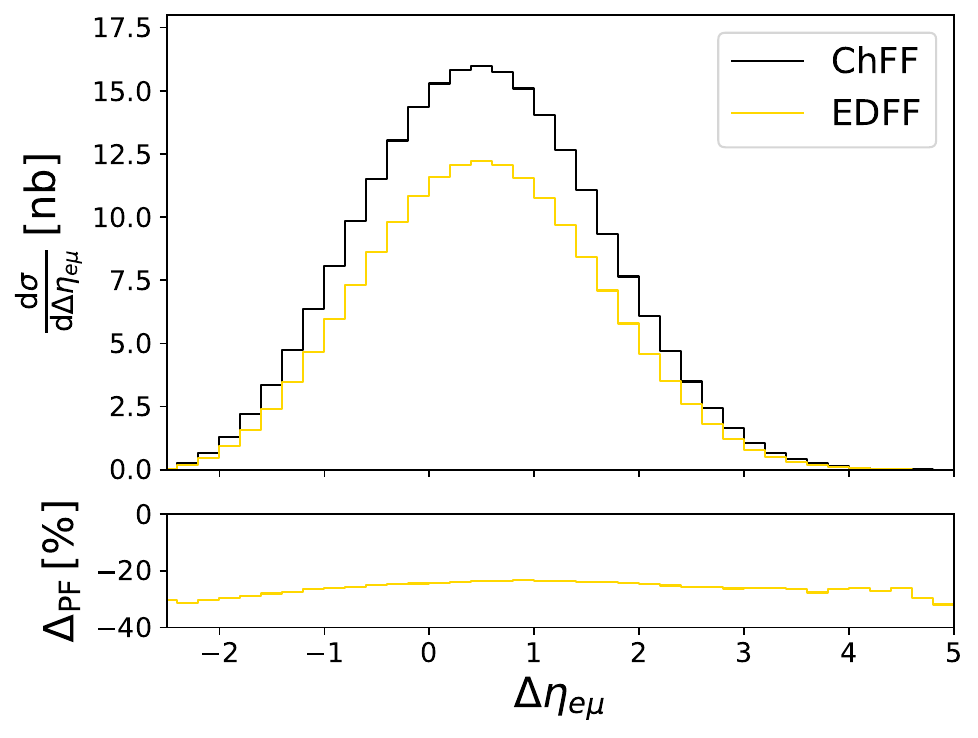}}
    \newline
    (c) \hspace{7.2cm} (d) \hspace{-1.3cm}
    }
    \caption{
    LO predictions for $\processmue$ induced by UPCs of two lead ions for:
    (a) the transverse momentum of the muon,
    (b) the pseudorapidity of the muon, 
    (c) the azimuthal angle difference between the muon and the positron, 
    and (d) the pseudorapidity difference between the muon and the positron.
    The photon flux is parametrized using the ChFF (black) and the EDFF (gold) of the heavy ions.
    The lower panels show the relative difference
    $\Delta_{\mathrm{PF}} = \sigma^\LO_{\mathrm{EDFF}}/\sigma^\LO_{\mathrm{ChFF}} - 1$.}
    \label{fig:PF_effects}
\end{figure}

In this section, the uncertainty arising from the parametrization of the photon flux is estimated by comparing LO results obtained using the EDFF and the ChFF approaches for the parametrization of the photon flux, see the second row in Table~\ref{tbl:PF_effects} and Fig.~\ref{fig:PF_effects}.
\begin{table}
  \centering{
    \begin{tabular}{c|c|c|c}
                                            \hl   & ChFF        & EDFF        & $\Delta_{\mathrm{PF}} [\%]$ \\
      \hline \hline
      $\sigma^{\LO} [\text{nb}]$            \hl   & $45.87(1)$  & $34.61(1)$  & $-24.56(1)$                 \\
      \hline
      $\sigma^{\LO}_{\mu\mu} [\mu\text{b}]$ \hl   & $57.24(2)$  & $45.64(1)$  & $-20.28(4)$                 \\ 
      \hline
      $\mathcal{O}^{\mathrm{LO}} \cdot 10^4 $ \hl & $8.013(3)$  & $7.583(3)$  & \wn$-5.38(5)$                  
    \end{tabular}
  }
  \caption{
  In the second and third rows, the LO predictions for the fiducial cross sections for $\processmue$ and for $\gamma\gamma\to\mu^+\mu^-$ induced by UPCs of two lead ions are provided.
  In the last row, the prediction for the normalized fiducial cross section,
as defined in (\ref{eq:normalized_sigma}), are presented.
  The second and third columns give the predictions using the ChFF and the EDFF to parametrize the photon flux, respectively.
  The relative difference
  $\Delta_{\mathrm{PF}} = \sigma^\LO_{\mathrm{EDFF}}/\sigma^\LO_{\mathrm{ChFF}} - 1$ is provided in the last column. 
  }
  \label{tbl:PF_effects}
\end{table}
We find that the prediction using the EDFF underestimates the result for the fiducial cross section by about $25\,\%$, assuming that the ChFF provides the more solid prediction. 
At the level of differential distributions, this effect is even more pronounced in phase-space regions with large muon transverse momentum, large muon pseudorapidity, and large pseudorapidity difference, reaching differences of $30\,\%$ and almost $40\,\%$ between the two predictions.
Similar effects have been already observed in Refs.~\cite{Dyndal:2020yen,Shao:2022cly}. 

In order to illustrate how this uncertainty can be reduced, we follow the strategy outlined in Sec.\,4 of Ref.\,\cite{Dyndal:2020yen}. 
This strategy relies on the ratios of cross sections and differential distributions to $\mu$-pair production,
\begin{align} \label{eq:normalized_sigma}
  \mathcal{O}_i \equiv \frac{\sigma^{\LO}_{i}}{\sigma^{\LO}_{\mu\mu,i}},
  \qquad
  \frac{\mathrm{d}\mathcal{O}_i}{\mathrm{d} X} \equiv \frac{1}{\sigma^{\LO}_{\mu\mu,i}} \frac{\mathrm{d} \sigma^{\LO}_{i}}{\mathrm{d} X},
\end{align}
with $i =$ EDFF, ChFF, and where $X$ stands for the corresponding differential variable. 
The LO prediction for the fiducial cross section for $\gamma\gamma\to\mu^+\mu^-$ induced by UPCs of two lead ions, $\sigma_{\mu\mu}^{\LO}$, are given in the third row of Table~\ref{tbl:PF_effects}. 
The fiducial phase space for di-muon production is defined by $p_{\text{T},\mu^{\pm}} > 4\,\GeV$ and $|\eta_{\mu^{\pm}}| > 2.5$. 
Note that, as in the predictions for $\processmue$, the LO fiducial cross section for di-muon production obtained using the EDFF for the parametrization of the photon flux differs from the one obtained using the ChFF by approximately $20\,\%$.

\begin{figure}
    \centering{
    \raisebox{0pt}{\includegraphics[width=0.5\columnwidth]{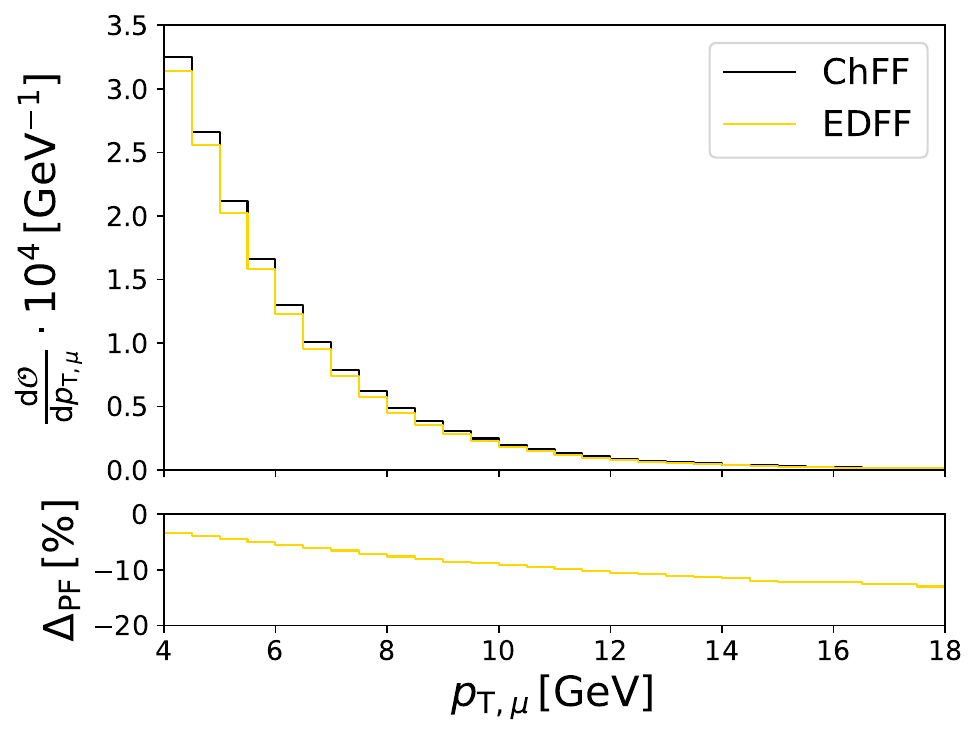}}
    \hspace{-0.3cm}
    \raisebox{0pt}{\includegraphics[width=0.5\columnwidth]{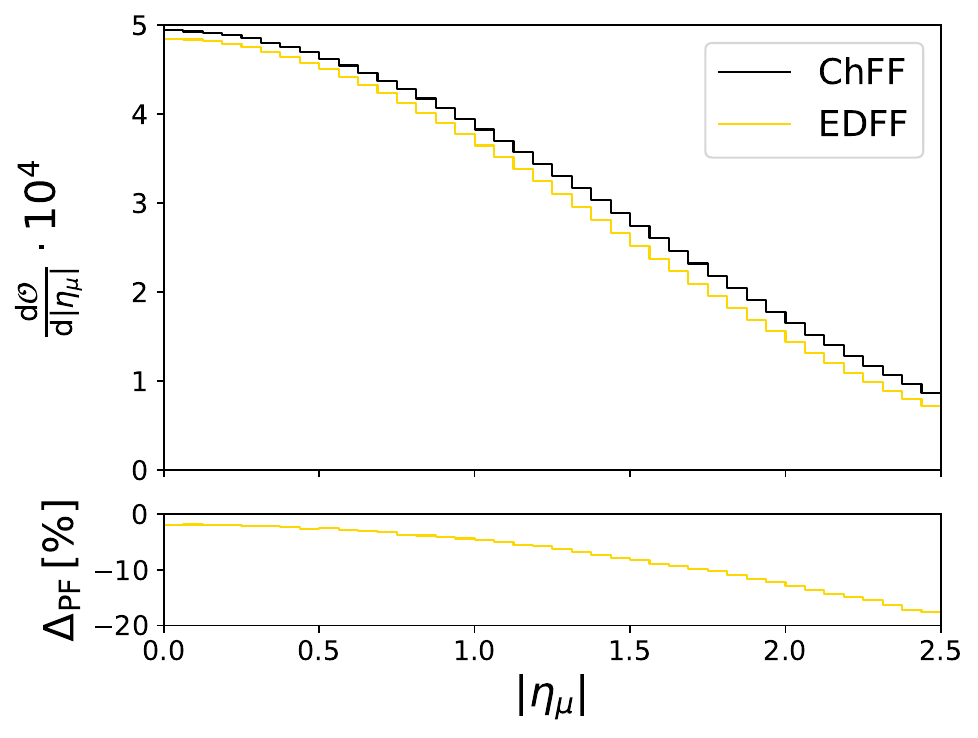}}
    \newline
    (a) \hspace{7.3cm} (b) \hspace{-5cm}
    \newline
    \raisebox{0pt}{\includegraphics[width=0.5\columnwidth]{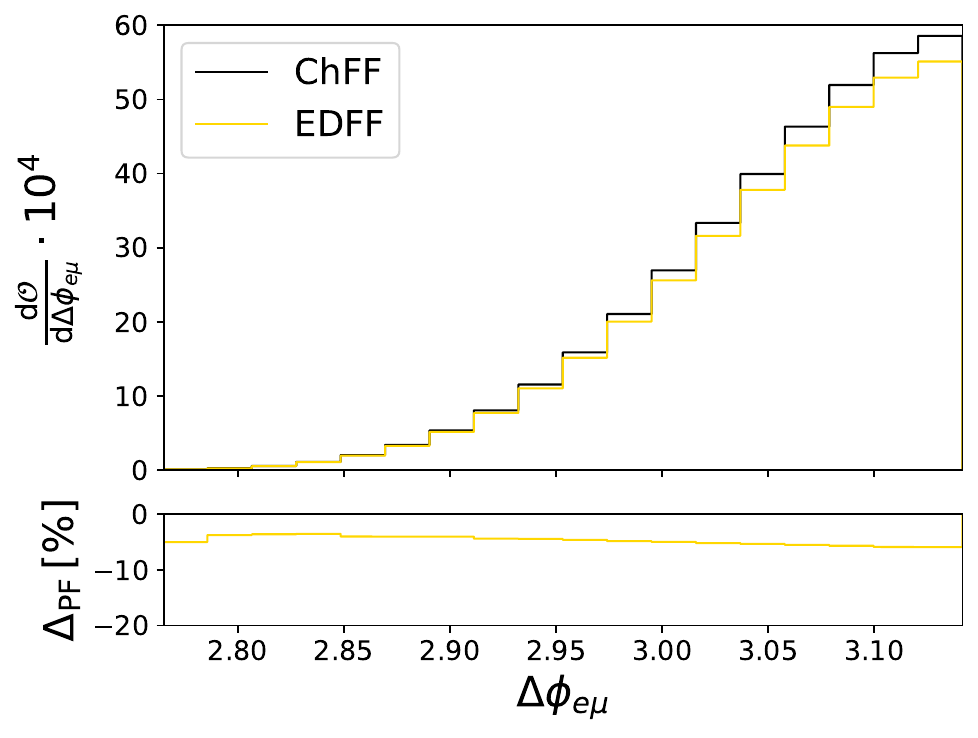}}
    \hspace{-0.3cm}
    \raisebox{0pt}{\includegraphics[width=0.5\columnwidth]{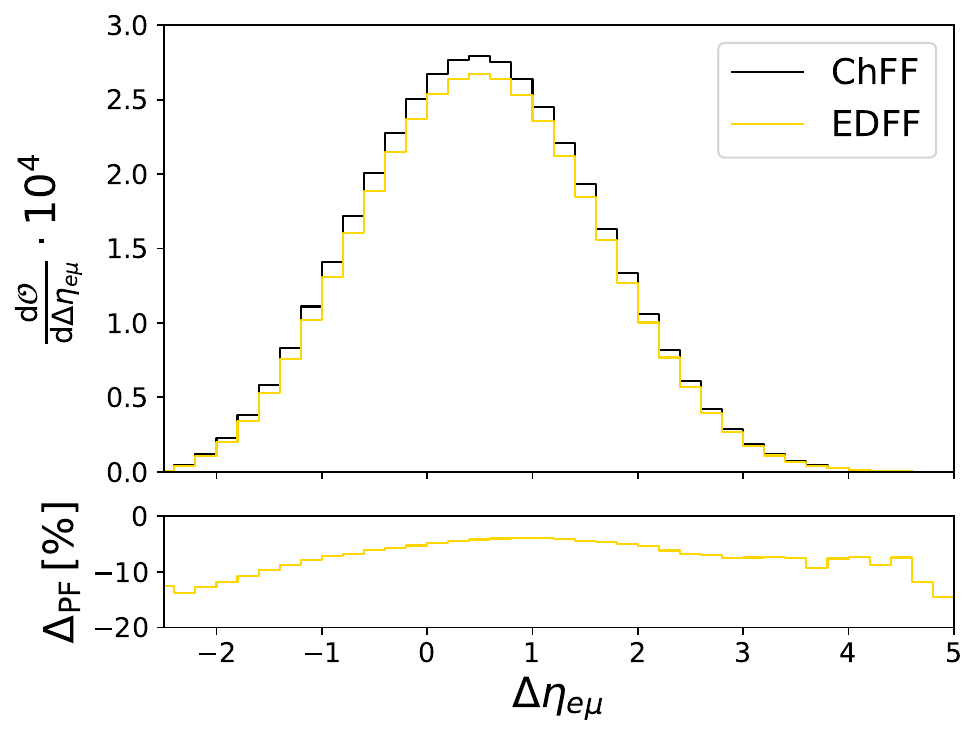}}
    \newline
    (c) \hspace{7.2cm} (d) \hspace{-1.3cm}
    }
    \caption{
    Same as Fig.~\ref{fig:PF_effects}, but for the LO predictions normalized by the LO prediction for the fiducial cross section of $\gamma\gamma\to\mu^+\mu^-$ induced by UPCs of two lead ions [see Eq.~\eqref{eq:normalized_sigma}].}
    \label{fig:O_PF_effects}
\end{figure}

The uncertainty arising from the parametrization of the photon flux partially cancels for the normalized observables, see the last row in Table~\ref{tbl:PF_effects} and Fig.~\ref{fig:O_PF_effects}.
The discrepancy between the two predictions is reduced to approximately $5\,\%$ for the normalized fiducial cross section, reaching up to $20\,\%$ in the less populated regions of the phase space.

\section{Conclusion}
\label{sec:conclusion}

In this article, we have presented a detailed study of theoretical predictions for the production of a 
pair of leptonically decaying $\tau$-leptons in ultraperipheral collisions of lead ions. 
The considered process $\processmue$ plays a central role in the precise determination of the anomalous magnetic moment of the $\tau$-lepton in heavy-ion runs of the LHC.
After the first observation of such process by ATLAS with the final state where both $\tau$-leptons decay leptonically, the present work provides the first SM predictions including spin correlations between the $\tau$-leptons, NLO EW corrections, and the full dependence on the masses of the final-state leptons.
The predictions make use of a state-of-the-art parametrization of the photon flux in UPCs using the charge form factor of the heavy ions.

The importance of spin-correlation effects is studied by comparing the LO predictions using a naive NWA and an improved version of the NWA which includes spin-correlation effects. 
It is shown that neglecting spin correlations between the $\tau$-leptons leads to an underestimate of the cross sections by about $5\,\%$.
Moreover, it is shown that while the massless approximation is valid for the positron, the error made by neglecting the mass of the muon is about $1\%$.

The aforementioned effects are of the same order of magnitude as the NLO EW correction, 
which is found to be negative and about $1.2\,\%$ of the LO contribution.
Thus, including spin correlations between the $\tau$-leptons and 
keeping the dependence on the masses of the final-state leptons is mandatory in precise predictions for the process under consideration. 

The largest contributions to the NLO EW correction come from the corrections to the $\tau$-lepton decays, which are about $-0.8\,\%$, while corrections to $\tau$-pair production are positive, around $0.4\,\%$. 
The contributions from closed fermion loops and the corrections due to the exchange of weak-gauge bosons in the $\tau$-pair production are found to be phenomenologically negligible
for an appropriately chosen (mixed) EW input-parameter scheme.
The dominant corrections are given by QED corrections to the $\tau$-pair production and 
the bosonic EW corrections to the leptonic $\tau$-decays.

Additionally, the impact of a non-inclusive treatment of collinear radiation off the muon on the NLO EW corrections has been analyzed for both massive muons and positrons, as well as in the massless limit. 
It has been observed that for bare muons, the uncancelled terms proportional to $\alpha\ln{m_\mu}$ do not lead to a pronounced collinear enhancement, as the muon mass is not much smaller than the 
probed energy range.
Applying a muon--photon recombination for collinear photon emission induces a redistribution of events in the NLO EW corrections to the transverse momentum of the muon.
This effect leads to an upshift in the prediction for the fiducial cross section, since the photon recombination lifts some muons into the fiducial region of the phase space.

Finally, the effect of different parametrizations of the photon flux is analyzed, showing that it is the major source of uncertainties. 
Consequently, further  improvements in this direction are necessary. 
Here, we confirm an earlier finding
that by considering observables based on ratios, \eg with respect to the prediction for $\gamma\gamma\to\mu^+\mu^-$, this uncertainty can be significantly reduced.  

In summary, we have provided a new state-of-the-art calculation for $\processmue$ induced by UPCs of lead ions that considerably improves on the precision of theory predictions and thus can contribute to a precise determination of the anomalous magnetic moment of the $\tau$-lepton in heavy-ion collisions at the LHC in the future.

\section*{Acknowledgements}

The authors would like to thank Hua-Sheng Shao for fruitful discussions on the parametrization of the photon flux and the use of \gammaUPC.
We thank Yann Stoll for providing checks on the matching between the leading-order matrix elements for $\tau$-pair production and $\tau$-decays.
Additionally, we are grateful to 
Kartik Bhide, Valerie Lang, and Markus Schumacher
for sharing their experimental insights.
Finally, the authors acknowledges support by the German Research Foundation (DFG) through the Research Training Group RTG2044 and through grant no INST 39/963-1 FUGG (bwForCluster NEMO) as well as the state of Baden-Württemberg through bwHPC.

\appendix
\section*{Appendix}

\section{Phase-space factorization within the narrow-width approximation}
\label{app:NWA_PS}

As discussed in Sec.~\ref{sec:hard_process}, the NWA is employed to evaluate matrix elements of the hard process.
Therefore, the only phase-space region that contributes to the cross section of the hard process is the one where the $\tau$-leptons are on-shell.
In this appendix, we show how this restriction affects the phase-space 
parametrizations at LO and NLO.

In the kinematical limit of the NWA, the LO contribution to the cross section for $\processmue$ is given by
\begin{align} \label{eq:sigma_LO}
  \hat{\sigma}^\LO 
  = \frac{1}{F} \int \mathrm{d} \Phi_6 \, \overline{|\M^{(0)}|^2}
  \,\, \underset{\mathrm{NWA}}{\longrightarrow} \,\,
    \frac{1}{F} \bigg(\frac{\pi}{m_{\tau}\Gamma_{\tau}}\bigg)^2 
    \int \mathrm{d} \Phi_6 \, 
    \overline{\big|\widetilde{\M}^{(0)}\big|^2} \,
    \delta(p_{\tau}^2 - m_{\tau}^2)
    \delta(\bar{p}_{\tau}^2 - m_{\tau}^2),
\end{align}
where $F = 2s_{\gamma\gamma}$ is the flux factor for the initial-state photons and 
$\mathrm{d}\Phi_6$ is the differential phase-space volume for the $6$-particle final state.
The square of the LO matrix element $\M^{(0)}$ of the full off-shell process is approximated using the NWA, as indicated in Eq.~(\ref{eq:ME2}). 

To parametrize the integral over the $6$-particle phase space in terms of the momenta of the final-state particles, we use the general expression for an $N$-particle phase space,
\begin{align} \label{eq:PS_N}
  \int \mathrm{d} \Phi_N (P) 
  = \bigg(\prod_{i=1}^N
    \Big[ 
      \int
      \frac{\mathrm{d}^4p_i}{(2\pi)^4} 
      (2\pi)\delta_+(p_i^2 - m_i^2)
    \Big]
    \bigg)
    (2\pi)^4 
    \delta^{(4)}\Big(P - \sum_{i=1}^N p_i\Big) ,
\end{align}
where $P = P_1 + P_2$ is the total incoming momentum, $P_{1,2}$ are the momenta of the initial-state photons, and $p_i$ and $m_i$ denote the momentum and the mass of the $i$-th final-state particle, respectively. 
We have introduced the notation $\delta_+(p^2 - m^2) \equiv \delta(p^2 - m^2)\theta(p^0)$.
In particular, for the $6$-particle final state, the phase-space integral can be written as
\begin{align} 
  \int \mathrm{d} \Phi_6 (P)
  = &\bigg(
    \prod_{i \in S_{\mathrm{D}}}
    \Big[ 
      \int
      \frac{\mathrm{d}^4p_i}{(2\pi)^4}
      (2\pi)\delta_+(p_i^2 - m_i^2) 
    \Big] 
    \bigg) 
    \bigg(
    \prod_{j \in S_{\,\overline{\mathrm{D}}}}
    \Big[ 
      \int
      \frac{\mathrm{d}^4p_j}{(2\pi)^4} 
      (2\pi)\delta_+(p_j^2 - m_j^2)
    \Big]
    \bigg) 
  \notag \\
  &\cdot
    (2\pi)^4
    \delta^{(4)}\Big(P - \sum_{i \in S} p_i \Big),
\end{align}
where we have split the set of final-state particles $S = S_\mathrm{D} \cup S_{\,\overline{\mathrm{D}}}$ into two subsets $S_\mathrm{D}$ and $S_{\,\overline{\mathrm{D}}}$, each of which formed by the final-state particles originating from the $\tau^-$- and $\tau^+$-decay, respectively.

In order to express the phase-space volume of the final state in terms of the phase-space volumes of the final states of the $\tau$-decays, we employ the identity
\begin{align} \label{eq:delta_id}
  \delta^{(4)}(P-\sum_{i\in S} p_i)
  =& \int \mathrm{d}^4p_\tau 
    \int \mathrm{d}M^2 
    \int \mathrm{d}^4\bar{p}_\tau 
    \int \mathrm{d}\bar{M}^2 \,
    \delta^{(4)}\Big(P - p_\tau - \bar{p}_\tau \Big)
  \notag \\
  &\cdot
    \delta_+(p_\tau^2 - M^2)\delta_+(\bar{p}_\tau^2 - \bar{M}^2)
    \delta^{(4)}\Big(p_\tau - \sum_{i\in S_{\mathrm{D}}} p_i \Big)
    \delta^{(4)}\Big(\bar{p}_\tau - \sum_{j\in S_{\,\overline{\mathrm{D}}}} p_j \Big)
    ,
\end{align}
where the integral on the right-hand side runs over all possible values for the auxiliary momenta $p_\tau$ and $\bar{p}_\tau$, and over the squared invariant masses $M^2$ and $\bar{M}^2$, whose integration boundaries are given by 
$\Big(\sum_{i\in S_{\mathrm{D}}} m_i \Big)^2 \leq  M^2 \leq \Big(\sqrt{s} - \sum_{i\notin X_{\mathrm{D}}} m_i \Big)^2$ and 
$\Big(\sum_{i\in S_{\,\overline{\mathrm{D}}}} m_i \Big)^2 \leq  \bar{M}^2 \leq \left(\sqrt{s} - M \right)^2$.
Thus, the integral over the $6$-particle final state can be expressed as
\begin{align} \label{eq:6_PS} 
  \int \mathrm{d} \Phi_6 (P)
  = &\frac{1}{(2\pi)^2} 
    \int \mathrm{d}M^2 
    \int \mathrm{d}\bar{M}^2 
    \int \mathrm{d}\Phi_{S_\mathrm{D}}(p_\tau) 
    \int \mathrm{d}\Phi_{S_{\,\overline{\mathrm{D}}}}(\bar{p}_\tau)
    \Big[\int\frac{\mathrm{d}^4p_\tau}{(2\pi)^4}(2\pi)\delta_+(p_\tau^2 - M^2)\Big]
  \notag \\
  & \cdot 
    \Big[\int\frac{\mathrm{d}^4\bar{p}_\tau}{(2\pi)^4}(2\pi)\delta_+(\bar{p}_\tau^2-\bar{M}^2)\Big]
    (2\pi)^4 \delta^{(4)}\Big(P - p_\tau - \bar{p}_\tau\Big),
\end{align}
where $\Phi_{S_{\mathrm{D}}}$ and $\Phi_{S_{\,\overline{\mathrm{D}}}}$ represent the phase spaces of the particles produced in the $\tau^-$- and $\tau^+$-decays, respectively. 

Note that the auxiliary momenta are chosen according to the momenta of the resonant $\tau$-leptons. 
Therefore, the $\delta$-functions arising from the narrow-width limit in Eq.~(\ref{eq:sigma_LO}) can replace the ones appearing in the $\delta_+$-functions in Eq.~(\ref{eq:6_PS}), so that the LO contribution to the cross section of the hard process can be calculated within the NWA as
\begin{align} 
  \hat{\sigma}^\LO 
  &= \frac{1}{F} \bigg(\frac{\pi}{m_{\tau}\Gamma_{\tau}}\bigg)^2 
     \int \mathrm{d} \Phi_6 \, 
     \overline{\big|\widetilde{\M}^{(0)}\big|^2} \,
     \delta(p_{\tau}^2 - m_{\tau}^2)
     \delta(\bar{p}_{\tau}^2 - m_{\tau}^2)
  \notag \\ 
  &= \frac{1}{F} \bigg(\frac{\pi}{m_{\tau}\Gamma_{\tau}}\bigg)^2 
     \frac{1}{(2\pi)^2}
     \int \mathrm{d}\Phi_{S_\mathrm{P}}(P) 
     \int \mathrm{d}\Phi_{S_\mathrm{D}}(p_\tau) 
     \int \mathrm{d}\Phi_{S_{\,\overline{\mathrm{D}}}}(\bar{p}_\tau)  \,
     \overline{\big|\widetilde{\M}^{(0)}\big|^2}
  \notag \\
  &\mwm \, \cdot \int\mathrm{d}M^2 \int\mathrm{d}\bar{M}^2 \, 
     \delta(p_{\tau}^2 - M^2)
     \delta(\bar{p}_{\tau}^2 - \bar{M}^2)
  \notag \\
  &= \frac{1}{F} \bigg(\frac{\pi}{m_{\tau}\Gamma_{\tau}}\bigg)^2 
     \frac{1}{(2\pi)^2}
     \int \mathrm{d}\Phi_{S_\mathrm{P}}(P) 
     \int \mathrm{d}\Phi_{S_\mathrm{D}}(p_\tau) 
     \int \mathrm{d}\Phi_{S_{\,\overline{\mathrm{D}}}}(\bar{p}_\tau) \, 
     \overline{\big|\widetilde{\M}^{(0)}\big|^2},
\end{align}
where, in the last step, it has been used that all the dependence on $M$ and $\bar{M}$ is on the $\delta$-functions and, thus, the integrals over $M^2$ and $\bar{M}^2$ integrate to one. 
At LO, the particle subset $S_\mathrm{P}$ is formed by the resonant $\tau$-leptons, and $\Phi_{S_\mathrm{P}}$ is the corresponding phase space.

The NLO contributions to the cross section consist of the virtual and real contributions $\Delta\hat{\sigma}^{(1)}$ and $\Delta\hat{\sigma}^{(\gamma)}$, respectively,
\begin{align}
  \hat{\sigma}^\NLO = \hat{\sigma}^\LO + \Delta\hat{\sigma}^\NLO,
  \quad \quad
  \Delta\hat{\sigma}^\NLO = \Delta\hat{\sigma}^{(1)} + \Delta\hat{\sigma}^{(\gamma)},
\end{align}
with
\begin{align} \label{eq:sigma_NLO}
  &\Delta\hat{\sigma}^{(1)} 
  = \frac{1}{F} \int \mathrm{d} \Phi_6 \, \overline{|\delta\M^{(1)}|^2}
  \,\, \underset{\mathrm{NWA}}{\longrightarrow}\, 
    \frac{1}{F} \bigg(\frac{\pi}{m_{\tau}\Gamma_{\tau}}\bigg)^2 
    \int \mathrm{d} \Phi_6 \, 
    \overline{\big|\delta\widetilde{\M}^{(1)}\big|^2} \,
    \delta(p_{\tau}^2 - m_{\tau}^2)
    \delta(\bar{p}_{\tau}^2 - m_{\tau}^2), 
    \\
  &\Delta\hat{\sigma}^{(\gamma)} 
  = \frac{1}{F} \int \mathrm{d} \Phi_7 \, \overline{|\M^{(\gamma)}|^2}  
  \,\, \underset{\mathrm{NWA}}{\longrightarrow} \,
    \frac{1}{F} \bigg(\frac{\pi}{m_{\tau}\Gamma_{\tau}}\bigg)^2 
    \int \mathrm{d} \Phi_7 \, 
    \overline{\big|\widetilde{\M}^{(\gamma)}\big|^2} \,
    \delta(p_{\tau}^2 - m_{\tau}^2)
    \delta(\bar{p}_{\tau}^2 - m_{\tau}^2),
\end{align}
where the one-loop corrections 
$\overline{|\delta\M^{(1)}|^2}$ and the squared real-emission amplitude 
$\overline{|\M^{(\gamma)}|^2}$
are approximated using the NWA, as indicated in Eq.~(\ref{eq:ME2}). 
Furthermore, using the break-up defined in Eqs.~(\ref{eq:ME2_1L}) and~(\ref{eq:ME2_RE}), the corrections can be written as
\begin{align} 
  \label{eq:sigma_NLOV_NWA}
  &\Delta\hat{\sigma}^{(1)} 
  = \Delta\hat{\sigma}^{(1,\mathrm{P})} 
    + \Delta\hat{\sigma}^{(1,\mathrm{D})} 
    + \Delta\hat{\sigma}^{(1,\overline{\mathrm{D}})},
  \\
  \label{eq:sigma_NLOR_NWA}
  &\Delta\hat{\sigma}^{(\gamma)} 
  = \Delta\hat{\sigma}^{(\gamma,\mathrm{P})} 
    + \Delta\hat{\sigma}^{(\gamma,\mathrm{D})} 
    + \Delta\hat{\sigma}^{(\gamma,\overline{\mathrm{D}})},
\end{align}
with 
\begin{align} 
  &\Delta\hat{\sigma}^{(1,X)} 
  = \frac{1}{F} \bigg(\frac{\pi}{m_{\tau}\Gamma_{\tau}}\bigg)^2 
    \int \mathrm{d} \Phi_6 \, 
    \overline{\big|\delta\widetilde{\M}^{(1,X)}\big|^2} \,
    \delta(p_{\tau}^2 - m_{\tau}^2)
    \delta(\bar{p}_{\tau}^2 - m_{\tau}^2),
  \\
  &\Delta\hat{\sigma}^{(\gamma,X)} 
  = \frac{1}{F} \bigg(\frac{\pi}{m_{\tau}\Gamma_{\tau}}\bigg)^2 
    \int \mathrm{d} \Phi_7 \, 
    \overline{\big|\widetilde{\M}^{(\gamma,X)}\big|^2} \,
    \delta(p_{\tau}^2 - m_{\tau}^2)
    \delta(\bar{p}_{\tau}^2 - m_{\tau}^2).
\end{align}
Since the one-loop corrections and the LO contribution have the same final state, the same phase-space factorization can be applied in both of them. Thus,
\begin{align} 
  \Delta\hat{\sigma}^{(1,X)} 
   = \frac{1}{F} \bigg(\frac{\pi}{m_{\tau}\Gamma_{\tau}}\bigg)^2 
     \frac{1}{(2\pi)^2}
     \int \mathrm{d}\Phi_{S_\mathrm{P}}(P) 
     \int \mathrm{d}\Phi_{S_\mathrm{D}}(p_\tau) 
     \int \mathrm{d}\Phi_{S_{\,\overline{\mathrm{D}}}}(\bar{p}_\tau) \,  
     \overline{\big|\delta\widetilde{\M}^{(1,X)}\big|^2}.
\end{align}
However, the presence of an additional photon in the final state of real-emission corrections requires some modification of the phase-space parametrization.
Specifically, the extra photon is assigned to the subset $S_{\mathrm{P}}$, $S_{\mathrm{D}}$, or $S_{\,\overline{\mathrm{D}}}$, depending on whether it is emitted in $\tau$-pair production, $\tau^-$-decay, or $\tau^+$-decay.
This assignment is indicated by $S^\gamma_{\mathrm{P}}$, $S^\gamma_{\mathrm{D}}$, or $S^\gamma_{\,\overline{\mathrm{D}}}$, respectively.
Therefore, the real-emission corrections are evaluated as 
\begin{align} 
  \Delta\hat{\sigma}^{(\gamma,\mathrm{P})} 
  = \frac{1}{F} \bigg(\frac{\pi}{m_{\tau}\Gamma_{\tau}}\bigg)^2 
    \frac{1}{(2\pi)^2}
    \int \mathrm{d}\Phi_{S^\gamma_\mathrm{P}}(P) 
    \int \mathrm{d}\Phi_{S_\mathrm{D}}(p_\tau) 
    \int \mathrm{d}\Phi_{S_{\,\overline{\mathrm{D}}}}(\bar{p}_\tau) \,
    \overline{\big|\widetilde{\M}^{(\gamma,\mathrm{P})}\big|^2},
  \notag \\
  \Delta\hat{\sigma}^{(\gamma,\mathrm{D})} 
  = \frac{1}{F} \bigg(\frac{\pi}{m_{\tau}\Gamma_{\tau}}\bigg)^2 
    \frac{1}{(2\pi)^2}
    \int \mathrm{d}\Phi_{S_\mathrm{P}}(P) 
    \int \mathrm{d}\Phi_{S^\gamma_\mathrm{D}}(p_\tau) 
    \int \mathrm{d}\Phi_{S_{\,\overline{\mathrm{D}}}}(\bar{p}_\tau) \,  
    \overline{\big|\widetilde{\M}^{(\gamma,\mathrm{D})}\big|^2},
  \notag \\
  \Delta\hat{\sigma}^{(\gamma,\overline{\mathrm{D}})} 
  = \frac{1}{F} \bigg(\frac{\pi}{m_{\tau}\Gamma_{\tau}}\bigg)^2 
    \frac{1}{(2\pi)^2}
    \int \mathrm{d}\Phi_{S_\mathrm{P}}(P) 
    \int \mathrm{d}\Phi_{S_\mathrm{D}}(p_\tau) 
    \int \mathrm{d}\Phi_{S^\gamma_{\,\overline{\mathrm{D}}}}(\bar{p}_\tau) \, 
    \overline{\big|\widetilde{\M}^{(\gamma,\overline{\mathrm{D}})}\big|^2}.
\end{align}


\section{Construction of the dipole subtraction term for real corrections}
\label{app:NWA_DS}


Within the NWA, the squares of the real-emission amplitudes are evaluated in the region of phase space where the resonant propagators are evaluated on-shell. 
This evaluation gives rise to IR singularities that are not present in the full matrix element of the off-shell process. 
For real emission, these IR singularities result from the phase-space integration over the soft emission off resonant propagators.  
In the virtual corrections, these IR singularities arise from loops with photon exchange involving couplings to the resonances.
In this section, we describe the construction of a subtraction term for $\processmue$ in the NWA using the dipole subtraction formalism~\cite{Dittmaier:1999mb,Dittmaier:2008md,Basso:2015gca,Catani:1996vz,Catani:2002hc,Campbell:2004ch}.


For any subtraction method, the main idea is to introduce a subtraction term $\Delta\hat{\sigma}^{\text{sub}}$ to subtract the singular contributions from the real corrections $\Delta\hat{\sigma}^{(\gamma)}$, to integrate $\Delta\hat{\sigma}^{\text{sub}}$ over the singular degrees of freedom, and to add the result to the virtual correction $\Delta\hat{\sigma}^{(1)}$, so that the total NLO correction is 
left unchanged,
\begin{align} \label{eq:subtraction}
  \Delta\hat{\sigma}^{\NLO} 
  &= \Delta\hat{\sigma}^{(1)} 
   + \Delta\hat{\sigma}^{(\gamma)} 
   = \Delta\hat{\sigma}^{\NLO}_{\text{V}} 
   + \Delta\hat{\sigma}^{\NLO}_{\text{R}}, 
  \notag \\
  \Delta\hat{\sigma}^{\NLO}_{\text{V}} 
  &= \Delta\hat{\sigma}^{(1)}
   + \Delta\hat{\sigma}^{\text{sub}}, 
  \notag \\
  \Delta\hat{\sigma}^{\NLO}_{\text{R}} 
  &= \Delta\hat{\sigma}^{(\gamma)}
   - \Delta\hat{\sigma}^{\text{sub}},
\end{align}
with
\begin{align}
  \Delta\hat{\sigma}^{\text{sub}} 
  = \frac{1}{F}\int \mathrm{d} \Phi_{n+1} 
    \big|\M^{(\text{sub})}(\Phi_{n+1})\big|^2.
\end{align}
Thus, the NLO virtual and real corrections, 
denoted by
$\Delta\hat{\sigma}^{\NLO}_{\text{V}}$ and $\Delta\hat{\sigma}^{\NLO}_{\text{R}}$, respectively, are given by
\begin{align}
  \Delta\hat{\sigma}^{\NLO}_{\text{V}} 
  &= \frac{1}{F}\int \mathrm{d} \Phi_{n}  
    \Big[ \,
      \overline{\big|\M^{(1)}(\Phi_{n})\big|^2} \,
      + \int \mathrm{d} \Phi_1
      \big|\M^{(\text{sub})}(\Phi_{n+1})\big|^2 
    \Big], 
  \label{eq:NLOV} \\
  \Delta\hat{\sigma}^{\NLO}_{\text{R}} 
  &= \frac{1}{F}\int \mathrm{d} \Phi_{n+1} 
    \Big[ \,
      \overline{\big|\M^{(\gamma)}(\Phi_{n+1})\big|^2} \,
      - \big|\M^{(\text{sub})}(\Phi_{n+1})\big|^2 
      \Big].
  \label{eq:NLOR}
\end{align}
In Eq.(\ref{eq:NLOV}), the $(n+1)$-particle phase space $\mathrm{d} \Phi_{n+1}$ has been factorized into the $n$-particle phase space $\mathrm{d} \Phi_{n}$ associated with the kinematics of the LO process and the $1$-particle phase space $\mathrm{d} \Phi_{1}$ of the photon. 
The subtracted amplitude $\big|\M^{(\text{sub})}\big|^2$ is built in such a way that it asymptotically behaves like the squared real-emission matrix element in the IR (soft and collinear) limits~\cite{Denner:2019vbn,Dittmaier:1999mb,Dittmaier:2008md,Catani:1996vz,Catani:2002hc}, \ie
\begin{align}
  \big|\M^{(\text{sub})}\big|^2 \, \asymp{\text{IR limits}} \,
 \overline{\big|\M^{(\gamma)}\big|^2}.
\end{align}
Thus, the integrand in Eq.\,(\ref{eq:NLOR}) is integrable in the IR limits without regulators. 

Moreover, the subtraction term must be simple enough, so that it can be integrated analytically over the singular degrees of freedom using a regulator, \emph{e.g.}\ dimensional regularization or regularization by small mass parameters. 
In the absence of collinear singularities from initial-state radiation, the partially integrated amplitude, also called integrated subtraction term,
\begin{align}
  \big|\M^{(\text{int sub})}(\Phi_n)\big|^2 =  \int \mathrm{d}\Phi_1 \big|\M^{(\text{sub})}(\Phi_{n+1})\big|^2 ,
\end{align}
possesses the same IR poles as those arising in the one-loop amplitudes, but with opposite sign. 
Thus, the sum of the virtual correction and the integrated subtraction in Eq.~(\ref{eq:NLOV}) is free of IR poles.


Within the NWA, the subtraction term for $\processmue$ can be constructed from individual subtraction terms for each subprocess. 
This approach is in line with the decomposition of the NLO corrections given in Eqs.~(\ref{eq:sigma_NLOV_NWA}) and~(\ref{eq:sigma_NLOR_NWA}),
%
%
\begin{align}
  \Delta\hat{\sigma}^{\text{sub}} 
  = \Delta\hat{\sigma}^{\text{sub}}_{\mathrm{P}}
  + \Delta\hat{\sigma}^{\text{sub}}_{\mathrm{D}}
  + \Delta\hat{\sigma}^{\text{sub}}_{\overline{\mathrm{D}}},
\end{align}
with
\begin{align} 
  \Delta\hat{\sigma}^{(\text{sub},\mathrm{P})} 
  = \frac{1}{F} \bigg(\frac{\pi}{m_{\tau}\Gamma_{\tau}}\bigg)^2 
    \frac{1}{(2\pi)^2}
    \int \mathrm{d}\Phi_{S^\gamma_\mathrm{P}}(P) 
    \int \mathrm{d}\Phi_{S_\mathrm{D}}(p_\tau) 
    \int \mathrm{d}\Phi_{S_{\,\overline{\mathrm{D}}}}(\bar{p}_\tau) \,   
    \big|\widetilde{\M}^{(\text{sub},\mathrm{P})}(\Phi_{7})\big|^2 ,
  \notag \\
  \Delta\hat{\sigma}^{(\text{sub},\mathrm{D})} 
  = \frac{1}{F} \bigg(\frac{\pi}{m_{\tau}\Gamma_{\tau}}\bigg)^2 
    \frac{1}{(2\pi)^2}
    \int \mathrm{d}\Phi_{S_\mathrm{P}}(P) 
    \int \mathrm{d}\Phi_{S^\gamma_\mathrm{D}}(p_\tau) 
    \int \mathrm{d}\Phi_{S_{\,\overline{\mathrm{D}}}}(\bar{p}_\tau) \,   
    \big|\widetilde{\M}^{(\text{sub},\mathrm{D})}(\Phi_{7})\big|^2 ,
  \notag \\
  \Delta\hat{\sigma}^{(\text{sub},\overline{\mathrm{D}})} 
  = \frac{1}{F} \bigg(\frac{\pi}{m_{\tau}\Gamma_{\tau}}\bigg)^2 
    \frac{1}{(2\pi)^2}
    \int \mathrm{d}\Phi_{S_\mathrm{P}}(P) 
    \int \mathrm{d}\Phi_{S_\mathrm{D}}(p_\tau) 
    \int \mathrm{d}\Phi_{S^\gamma_{\,\overline{\mathrm{D}}}}(\bar{p}_\tau) \,   
    \big|\widetilde{\M}^{(\text{sub},\overline{\mathrm{D}})}(\Phi_{7})\big|^2 ,
\end{align}
where the squared amplitudes are evaluated in the narrow-width limit.


For each individual subtraction term $\Delta\hat{\sigma}^{(\text{sub},X)}$, we use the dipole subtraction formalism to build the subtraction amplitudes. 
Within this formalism, the subtraction amplitudes are constructed from so-called dipoles. 
Each dipole contains an emitter $f$ and a spectator $f'$. 
Only the kinematics of the emitters lead to IR singularities, the spectators are just needed to balance recoil effects and thus to guarantee momentum conservation. 
For each dipole, the subtraction term is a product of the squared LO matrix element and the 
dipole function $g_{ff',\xi}^{(\text{sub})}$ describing the radiation and depending only on the kinematics of emitter, spectator, and the radiated photon, see Refs.~\cite{Dittmaier:1999mb,Dittmaier:2008md,Basso:2015gca}.
All possible emitter--spectator configurations have to be summed over. 
The subtraction amplitude is explicitly given by
\begin{align} \label{eq:DS}
  \big|\widetilde{\M}^{(\text{sub},X)}(\Phi_{7})\big|^2  
  &= \sum_{f\neq f'}
    \big|\widetilde{\M}^{(\text{sub},X)}_{ff'}(\Phi_{7})\big|^2  ,
  \notag \\ 
  \big|\widetilde{\M}^{(\text{sub},X)}_{ff',\lambda_f}(\Phi_{7})\big|^2 
  &= -e^2Q_f\eta_fQ_{f'}\eta_{f'}
    g_{ff',\xi}^{(\text{sub})}(p_f, p_{f'},k)
    \big|\widetilde{\M}^{(\text{0})}_{\xi\lambda_f}(\tilde{\Phi}^{X}_{6,ff'})\big|^2 ,
\end{align}
where the sum runs over all possible emitter--spectator configurations. 
The relative charges of the emitter and spectator are 
denoted by $Q_f$ and $Q_{f'}$, respectively. 
The symbols $\eta_f$ and $\eta_{f'}$ indicate the charge flow of the emitter and spectator, respectively. They are equal to $+1$ for incoming particles or outgoing antiparticles and to $-1$ for incoming antiparticles or outgoing particles. 
Moreover, there is an implicit sum assumed over $\xi = \pm$, taking into account the possibility of the emitter keeping ($+$) or flipping ($-$) its polarization $\lambda_f$ during the emission of a photon with momentum $k$.
Note that the LO matrix element is evaluated on the reduced phase space $\tilde{\Phi}^{X}_{6,ff'}$.
For each dipole, the reduced phase space is the remaining phase space after factorizing the phase space associated with the radiated photon. 
The superscript $X=\mathrm{P}, \mathrm{D}, \overline{\mathrm{D}}$ in the reduced phase space indicates that only the matrix element associated to the subprocess $X$ is evaluated using the mapped momenta, 
\ie employing the naive version of the NWA,
\begin{align}
  \overline{\big|\widetilde{\M}^{(0)}_{\mathrm{NWA}} (\tilde{\Phi}^{\mathrm{P}}_{6,ij})\big|^2}
  = \overline{\big|\M^{(0)}_{\mathrm{P}}(\tilde{\Phi}_{S_{\mathrm{P}},ij})\big|^2} \;
    \overline{\big|\M^{(0)}_{\mathrm{D}}(\Phi_{S_{\mathrm{D}}})\big|^2} \;
    \overline{\big|\M^{(0)}_{\overline{\mathrm{D}}}(\Phi_{S_{\,\overline{\mathrm{D}}}})\big|^2},
  \notag \\
  \overline{\big|\widetilde{\M}^{(0)}_{\mathrm{NWA}} (\tilde{\Phi}^{\mathrm{D}}_{6,ia})\big|^2}
  = \overline{\big|\M^{(0)}_{\mathrm{P}}(\Phi_{S_{\mathrm{P}}})\big|^2} \;
    \overline{\big|\M^{(0)}_{\mathrm{D}}(\tilde{\Phi}_{S_{\mathrm{D}},ia})\big|^2} \;
    \overline{\big|\M^{(0)}_{\overline{\mathrm{D}}}(\Phi_{S_{\,\overline{\mathrm{D}}}})\big|^2},
  \notag \\
  \overline{\big|\widetilde{\M}^{(0)}_{\mathrm{NWA}} (\tilde{\Phi}^{\overline{\mathrm{D}}}_{6,ia})\big|^2}
  = \overline{\big|\M^{(0)}_{\mathrm{P}}(\Phi_{S_{\mathrm{P}}})\big|^2} \;
    \overline{\big|\M^{(0)}_{\mathrm{D}}(\Phi_{S_{\mathrm{D}}})\big|^2} \;
    \overline{\big|\M^{(0)}_{\overline{\mathrm{D}}}(\tilde{\Phi}_{S_{\,\overline{\mathrm{D}}},ia})\big|^2},
\end{align} 
and using the improved version of the NWA,
\begin{align}
  \widetilde{\M}_{\text{iNWA}}^{(0)} (\tilde{\Phi}^{\mathrm{P}}_{6,ij})
  = \sum_{\sigma,\bar{\sigma}}
    \M^{(0)}_{\text{P},\sigma\bar{\sigma}}(\tilde{\Phi}_{S_{\mathrm{P}},ij})
    \M^{(0)}_{\text{D},\sigma}(\Phi_{S_{\mathrm{D}}})
    \M^{(0)}_{\overline{\text{D}},\bar{\sigma}}(\Phi_{S_{\,\overline{\mathrm{D}}}}),
  \notag \\
  \widetilde{\M}_{\text{iNWA}}^{(0)} (\tilde{\Phi}^{\mathrm{D}}_{6,ia})
  = \sum_{\sigma,\bar{\sigma}}
    \M^{(0)}_{\text{P},\sigma\bar{\sigma}}(\Phi_{S_{\mathrm{P}}})
    \M^{(0)}_{\text{D},\sigma}(\tilde{\Phi}_{S_{\mathrm{D}},ia})
    \M^{(0)}_{\overline{\text{D}},\bar{\sigma}}(\Phi_{S_{\,\overline{\mathrm{D}}}}),
  \notag \\
  \widetilde{\M}_{\text{iNWA}}^{(0)} (\tilde{\Phi}^{\overline{\mathrm{D}}}_{6,ia})
  = \sum_{\sigma,\bar{\sigma}}
    \M^{(0)}_{\text{P},\sigma\bar{\sigma}}(\Phi_{S_{\mathrm{P}}})
    \M^{(0)}_{\text{D},\sigma}(\Phi_{S_{\mathrm{D}}})
    \M^{(0)}_{\overline{\text{D}},\bar{\sigma}}(\tilde{\Phi}_{S_{\,\overline{\mathrm{D}}},ia}),
\end{align} 
where we used the common notation $f=i,j,k\ldots$, if the emitter/spectator or the fermion is a final-state particle and $f=a,b,c\ldots$, if it is an initial-state particle.
The mappings
$\Phi_{S_\mathrm{P}}^\gamma \to \tilde{\Phi}_{S_\mathrm{P},ij}$, 
$\Phi_{S_\mathrm{D}}^\gamma \to \tilde{\Phi}_{S_\mathrm{D},ia}$, and 
$\Phi_{S_{\,\overline{\mathrm{D}}}}^\gamma \to \tilde{\Phi}_{S_{\,\overline{\mathrm{D}}},ia}$
are performed following the constructions given in Refs.~\cite{Dittmaier:1999mb,Dittmaier:2008md,Basso:2015gca}.

For each of the subprocesses, the emitter--spectator configurations contributing to the sum in Eq.~(\ref{eq:DS}) differs. 
In particular, for $\tau$-pair production, there are no charged particles in the initial state and, thus, only final-state emitter and final-state spectator configurations with one of the resonant $\tau$-leptons as emitter and the other as spectator contributes to the sum, \ie
\begin{align}
  \big|\widetilde{\M}^{(\text{sub},\mathrm{P})}(\Phi_{7})\big|^2 
  = 4\pi\alpha 
    \Big[ g^{(\text{sub})}_{\tau\bar{\tau}}(p_\tau, \bar{p}_{\tau}, k) \,
    \overline{\big|\widetilde{\M}^{(0)}(\tilde{\Phi}^{\mathrm{P}}_{6,\tau\bar{\tau}})\big|^2} \,
  + g^{(\text{sub})}_{\bar{\tau}\tau}(\bar{p}_{\tau}, p_\tau, k) \,
    \overline{\big|\widetilde{\M}^{(0)}(\tilde{\Phi}^{\mathrm{P}}_{6,\bar{\tau}\tau})\big|^2} \,
    \Big].
\end{align}
Note that since the polarization of the $\tau$-leptons is always summed over, the unpolarized dipole function 
$g_{i j}^{(\text{sub})} = g_{i j,+}^{(\text{sub})} + g_{i j,-}^{(\text{sub})}$ 
is used, and no polarization flip occurs in the evaluation of the LO matrix element.
Note also that the $\tau$-leptons are always considered massive.
The corresponding dipole functions $g_{ij,\xi}^{(\text{sub})}$ can be found in Ref.~\cite{Dittmaier:1999mb}.

For the decays of the $\tau$-leptons, as for every decay process, the mass of the initial-state particle determines the hard energy scale of the process.
Thus, it is never negligible, and there are no collinear singularities associated to initial-state radiation. 
This allows us to combine subtraction functions for configurations where the initial-state particle $a$ acts either as the emitter or as the spectator into a single subtraction function $d^{(\text{sub})}_{ia,\xi} = g^{(\text{sub})}_{ia,\xi} + g^{(\text{sub})}_{ai,\xi}$.
Since there is only one charged particle in the final state of each decay, there are no final-state emitter and final-state spectator contributions.
Therefore, the subtraction terms for the decays can be built as
\begin{align}
  &\big|\widetilde{\M}^{(\text{sub},\mathrm{D})}(\Phi_{7})\big|^2 
  = 4\pi\alpha\, d^{(\text{sub})}_{\mu\tau}(p_\mu, p_\tau, k) \,
     \overline{\big|\widetilde{\M}^{(0)}(\tilde{\Phi}^{\mathrm{D}}_{6,\mu\tau})\big|^2},
  \notag \\
  &\big|\widetilde{\M}^{(\text{sub},\overline{\mathrm{D}})}(\Phi_{7})\big|^2 
  = 4\pi\alpha\, d^{(\text{sub})}_{e\bar{\tau}}(p_e, \bar{p}_\tau, k) \,
     \overline{\big|\widetilde{\M}^{(0)}(\tilde{\Phi}^{\overline{\mathrm{D}}}_{6,e\bar{\tau}})\big|^2},
\end{align}
where $p_\mu$ and $p_e$ denote the muon and the positron momenta, respectively. 
The dipole functions 
$d_{ia}^{(\text{sub})} = d_{ia,+}^{(\text{sub})} + d_{ia,-}^{(\text{sub})}$
can be found in Ref.~\cite{Basso:2015gca}.
In the case that the massless limit is considered for the final-state charged leptons and the collinear radiation off leptons is not treated inclusively, particular care needs to be taken in the application of the subtraction procedure.
The corresponding modifications to the subtraction procedure to treat IR-non-safe configurations are described in Ref.~\cite{Dittmaier:2008md}.

\newpage

\bibliographystyle{utphys.bst}
\bibliography{aa-tautau}

\end{document}